\documentclass[twocolumn,amsmath,amssymb,superscriptaddress,prb,showkeys,showpacs,nofootinbib]{revtex4-2}
\usepackage{natbib}
\usepackage{graphicx}
\usepackage{dcolumn}
\usepackage{bm}
\usepackage[utf8]{inputenc}
\usepackage[T1]{fontenc}
\usepackage{mathptmx}
\usepackage{etoolbox}

\usepackage{easyReview}
\usepackage{microtype}
\usepackage{upgreek}
\usepackage{booktabs}
\usepackage{multirow}
\usepackage{tabularx}
\usepackage{xcolor}
\definecolor{link}{RGB}{57,106,177}
\definecolor{blue}{RGB}{0,0,0}
\usepackage{hyperref}
\hypersetup{
	colorlinks=true,
	linkcolor=link,
	citecolor=link,
	urlcolor=link
}
\usepackage{cleveref}
\usepackage{wasysym}
\newcommand{\ped}[1]{\ensuremath{_{\rm #1}}}
\newcommand{\apex}[1]{\ensuremath{^{\rm #1}}}
 
\makeatother
\begin{document}

\title{Spectroscopic studies of the superconducting gap in the 12442 family of iron-based compounds}

\author{Erik Piatti}
\affiliation{\mbox{Department of Applied Science and Technology, Politecnico di Torino, I-10129 Torino, Italy}}
\author{Daniele Torsello}
\author{Gianluca Ghigo}
\affiliation{\mbox{Department of Applied Science and Technology, Politecnico di Torino, I-10129 Torino, Italy}}
\affiliation{Istituto Nazionale di Fisica Nucleare, Sezione di Torino, I-10125 Torino, Italy}
\author{Dario Daghero}
\email{dario.daghero@polito.it}
\affiliation{\mbox{Department of Applied Science and Technology, Politecnico di Torino, I-10129 Torino, Italy}}

\begin{abstract}
The iron-based compounds of the so-called 12442 family are very peculiar in various respects. They originate from the intergrowth of 122 and 1111 building blocks, display a large in-plane vs. out-of-plane anisotropy, possess double layers of FeAs separated by insulating layers, and are generally very similar to double-layer cuprates. Moreover, they are stoichiometric superconductors because of an intrinsic hole doping. Establishing their superconducting properties, and in particular the symmetry of the order parameter, is thus particularly relevant in order to understand to what extent these compounds can be considered as the iron-based counterpart of cuprates. In this work we review the results of various techniques from the current literature and compare them with ours, obtained in Rb-12442 by combining point-contact Andreev-reflection spectroscopy and coplanar waveguide resonator measurements of the superfluid density. It turns out that the compound possesses at least two gaps, one of which is certainly nodal. The compatibility of this result with the theoretically allowed gap structures, as well as with the other results in literature, is discussed in detail.\\\\
Cite this article as: E. Piatti, D. Torsello, G. Ghigo, and D. Daghero. \textit{Low Temp. Phys.} \href{https://doi.org/10.1063/10.0019688}{\textbf{49}, 770--785 (2023)}.   
\end{abstract}


\maketitle

\section{Introduction: the 12442 iron-based superconductors}

\begin{figure}
    \centering
    \includegraphics[width=\linewidth]{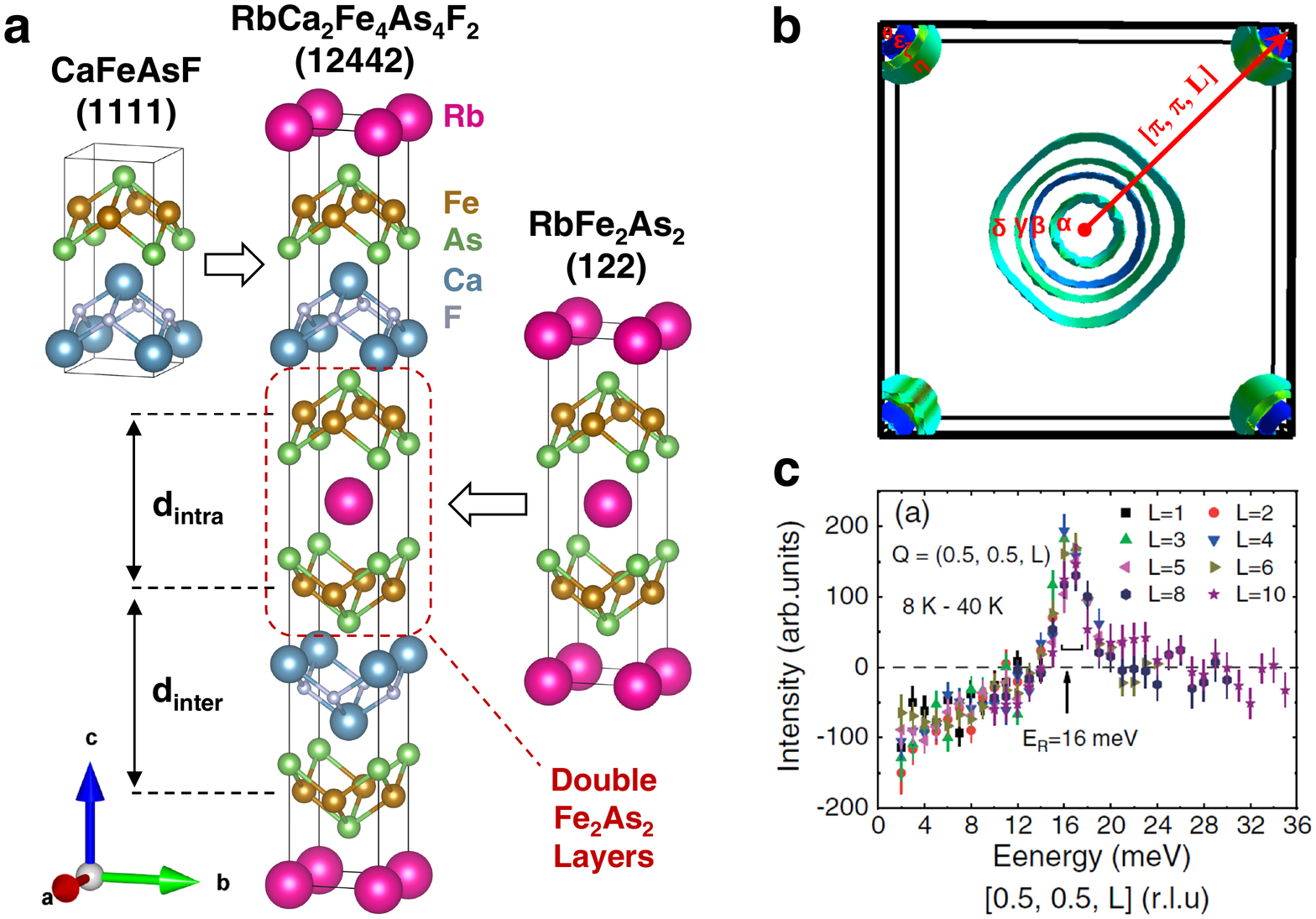}
    \caption{
    (a) Ball-and-stick model of the Rb-12442 structure (center) and of its building blocks, the 1111-type CaFeAsF (left) and 122-type RbFe$_2$As$_2$ (right) structures, as rendered by VESTA\,\cite{vesta}. The distance between the two adjacent FeAs layers belonging to the 122-type bilayer ($d\ped{intra}$), and the distance between the two neighboring FeAs layers belonging to one 122-type and one 1111-type bilayer ($d\ped{inter}$), are highlighted.
    (b) Top view of the Fermi surfaces of the 12442 family of iron-based compounds computed via Density Functional Theory in the 2-Fe Brillouin Zone. Reprinted with permission from Ref.\,\onlinecite{Ghosh2020}. Copyright 2020 by Elsevier.
    (c) Spin resonant peak detected by inelastic neutron scattering in K-12442 at the in-plane wavevector $Q=(0.5, 0.5)$ (in units of $2\pi/a$) of the 2-Fe Brillouin Zone. Reprinted with permission from Ref.\,\onlinecite{Hong2020PRL}. Copyright 2020 by the American Physical Society. The same wavevector is shown in panel (b) as the red arrow at $Q=(\pi, \pi)$ (in units of $1/a$).
    }
    \label{fig:structure}
\end{figure}

Among the various classes of iron-based superconductors, the most recently discovered is the so-called 12442 family\,\cite{Wang2016JACS, Wang2017SciChiMat, Wu2017PRM}. These compounds have a complicated structure, shown in Figure\,\ref{fig:structure}a, that results from the intergrowth of 122-type AFe$_2$As$_2$ (where A is an alkaline metal, i.e. K, Rb, or Cs), and 1111-type CaFeAsF or LnFeAsO\,\cite{Wu2017PRM} where Ln is a lanthanide (Gd, Tb, Dy etc). The resulting unit formula is thus ACa$_2$Fe$_4$As$_4$F$_2$ or ALn$_2$Fe$_4$As$_4$O$_2$, hence the name ``12442" of the family. Here we will focus on the first kind of compounds, thus excluding the oxyarsenides. The structure is body-centred tetragonal, belonging to the space group I4/mmm. The AFe$_2$As$_2$ building block is heavily hole-doped with 0.5 holes per Fe atom, while the CaFeAsF is undoped; as a result, the compound as a whole is intrinsically doped with 0.25 holes per Fe atom\,\cite{Wang2016EPL}. This has the important and very relevant consequence that these compounds do not show any trace of long-range antiferromagnetic spin-density wave (SDW) order. In some sense, if one imagines the generic phase diagram of iron-based compounds, their intrinsic doping places them far from the region in which the SDW order is progressively destroyed by doping or substitutions, in favour of the appearance of superconductivity. Indeed, these compounds are stoichiometric superconductors with a critical temperature that ranges between 28\,K and 33\,K depending on the alkaline metal A. 
The structure is such that two layers of FeAs, with alkaline metal sandwiched in between, are separated from the other two layers by an insulating layer of Ca$_2$Fe$_2$. This is a first structural similarity with the double-layer cuprates (for instance Bi$_2$Sr$_2$CaCu$_2$O$_{8+\delta}$ or YBa$_2$Cu$_3$O$_{6+\delta}$) that is unique among iron-based superconductors. 
Another consequence of the structure is the large anisotropy of the resistivity, $\gamma_{\rho}$\,\cite{Wang2019prb} and of the upper critical field, $\gamma_H$\,\cite{Yi2020NJP, Wang2020SciChi_pulsed} (which near $T_c$ is related to the anisotropy of the resistivity,\cite{LiuPRB2014, TanatarPRB2009} by $\gamma_{\rho} \sim \gamma_H^2$). The values of the anisotropy are higher than in all other iron-based compounds, with the exception of some electron-doped 1111 compounds, that feature thick spacer layers\,\cite{Wang2019prb}: in particular, for 11 and 122 families $\gamma_H \sim 3$. The value of the anistropy parameters in the 12442 compounds is rather similar to those of some cuprates. Magnetic torque measurements on K-12442 single crystals provided an almost temperature-independent anisotropy factor $\gamma_H \sim 16$\,\cite{Yu2019PRB}.

Iron-based compounds are known to have superconducting properties that strongly depend on tiny details of the structure, namely the height of the pnictogen atom (here As) above the FeAs plane (usually referred to as $h\ped{As}$) and the angle made by the two bonds between a Fe atom and the nearby As atoms, i.e. the As-Fe-As bond angle, usually referred to as $\alpha$. In particular, a fairly general rule relates the maximum of the critical temperature with a particular combination of these parameters, i.e. $h\ped{As} = 1.38 $ \AA ~and $\alpha = 109.5$°. In the case of 12442 compounds, FeAs planes are asymmetric (this is a feature that they share with the 1144 family), and therefore there are actually two different values of each parameter, depending on whether it is measured on the side of the FeAs plane toward the alkaline metal or on the side that faces the Ca$_2$F$_2$ plane. As pointed out in Ref.\,\onlinecite{Wang2017SciChiMat}, the experimental values of these parameters are different from those expected to maximize $T_c$. Surprisingly, the change of the alkaline metal from K to Rb and then to Cs, with increasing atomic size, makes the $T_c$ decrease but, at the same time, leads the compound closer to the optimal values of $h\ped{As}$ and $\alpha$. Clearly, the asymmetry of the FeAs layer makes a description in terms of these structural parameters simplistic or simply unreliable.

Another interesting parameter that seems to be related to the tuning of $T_c$ is the interlayer spacing. Indeed, the structure of the compound implies that there are two different distances between FeAs layers. One is the distance between the layers that are separated by only the alkaline atom and form a bilayer (intralayer distance, $d\ped{intra}$ in the following) and the other is the distance between neighbouring layers that actually belong to different bilayers (interlayer distance $d\ped{inter}$). Going from K to Rb and then to Cs, the intralayer distance increases while the interlayer distance decreases, and the critical temperature decreases as well. This somewhat suggests that interlayer distance is a critical parameter and that the farther the double layers are, the higher is $T_c$. This is interesting because the interlayer distance is also greater than the c-axis coherence length\,\cite{Yi2020NJP, Wang2019prb} which indicates that superconductivity has a quasi-2D character and the bilayers are decoupled. 

As in almost all the iron-based compounds, the Fermi surface (FS) is made of various sheets. Here, due to the large anisotropy, these sheets are almost perfect cylinders and do not show the warping which is observed, for example, in 122 compounds\,\cite{GonnelliSCIREP2016, Daghero2013LTP}. According to first-principle calculations\,\cite{Ghosh2020, Ishida2017PRB}, 12442 compounds present up to 8 hole-like sheets centred on the $\Gamma$ point of the Brillouin zone (two pairs are nearly degenerate) and other 4 electron-like sheets centred at the M points, at the corners of the Brillouin zone (see Figure\,\ref{fig:structure}b). Direct measurements of the FS structure carried out by means of angle-resolved photoemission spectroscopy (ARPES) in K-12442\,\cite{Wu2020PRB} essentially confirmed this picture, with three clearly separated hole-like cylinders at the zone centre (two of which actually consist of two nearly degenerate surfaces) and a small electron-like pocket at M, as shown in Figure\,\ref{fig:ARPES}a. The unusual disproportion between the size of the hole-like sheets and the electron-like ones is clearly due to the large intrinsic hole doping. As a consequence, there is no nesting wavevector able to connect the electron-like and hole-like bands, which is generally the condition under which a long-range stripe antiferromagnetic order can appear. Indeed, first-principle calculations indicate that the ground state would display a stripe antiferromagnetic order\,\cite{Wang2016EPL}, which is however suppressed by self-doping.

The absence of a long-range spin order, however, does not prevent the spin susceptibility from presenting a clear resonance at an energy $E\ped{R}$ that, in K-12442, is equal to 16\,meV\,\cite{Hong2020PRL} (see Figure\,\ref{fig:structure}c). Considering the critical temperature of the compound, the ratio $E\ped{R}/k\ped{B} T_c$ is about 5.5, a value which is similar to that of cuprates (5.8) but unprecedented in iron-based compounds. As a matter of fact, a generally rather well-obeyed rule in iron-based superconductor is that the energy of the spin resonance, which is believed to correspond to the energy of the mediating boson in the picture of spin-fluctuation-mediated pairing, is related to the critical temperature by $\Omega_b \simeq 4.65 k\ped{B} T_c$\,\cite{Paglione2010}. The direct determination of the characteristic energy of the electron-boson spectral function, obtained by analyzing point-contact spectroscopy spectra in iron-based superconductors of the 1111\,\cite{Daghero2011, DagheroPRB2020} and 122\,\cite{Tortello2010} families, always confirmed this picture.
The spin resonance, determined by inelastic neutron scattering experiment, is observed in correspondence of an in-plane wavevector $Q=(\pi, \pi)$ (red arrow in Figure\,\ref{fig:structure}b) which connects the center of the 2-Fe Brillouin zone $\Gamma$ with its corner M\,\cite{Hong2020PRL}. In most iron-based compounds (apart from the 11 chalcogenides), the spin-resonance wavevector is also the wavevector of the long-range antiferromagnetic order of the undoped compound. These evidences can be naturally explained within the models for spin-fluctuation-mediated superconductivity. It is not completely clear to what extent a ``quasi nesting model"\,\cite{Richard2011RoPP, Hirschfeld2011RoPP} can be applied to 12442 compounds, owing to the aforementioned disproportion between hole-like and electron-like FS sheets, or wether the local antiferromagnetic exchange model must be applied\,\cite{Richard2011RoPP, Hirschfeld2011RoPP}.  

\section{Gap symmetry in 12442 compounds}

\begin{figure*}
    \centering
    \includegraphics[width=\linewidth]{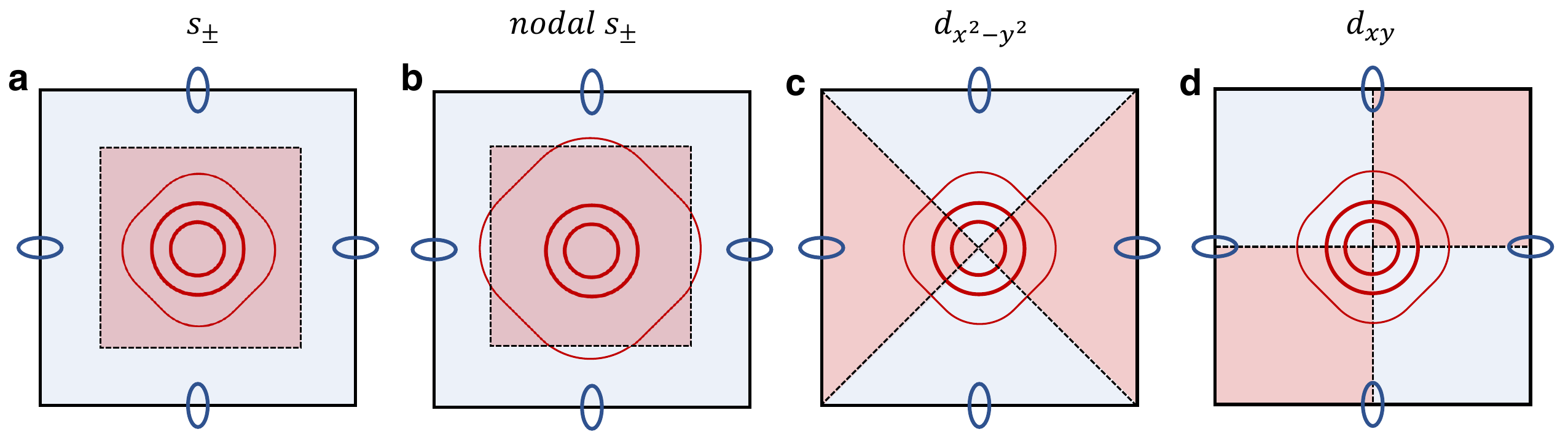}
    \caption{
    Sketches of the symmetries of the superconducting energy gaps considered in this work and of the Fermi surfaces (FS) of 12442 compounds represented in the 2D 1-Fe Brillouin zone. The large hole-like Fermi surface sheets at the zone center and the small electron pockets on the zone sides are shown. The regions shaded in blue/red correspond to different signs of the order parameter. Dashed lines highlight the nodal lines.
    (a) Nodeless $s_\pm$ symmetry. No gaps show nodal lines on any FS.
    (b) Nodal $s_\pm$ symmetry. The nodes are accidental, either on the hole-like FS (depicted here) or the electron-like FS (not shown).
    (c) Nodal $d_{x^2-y^2}$ symmetry. The nodes on the hole-like Fermi surfaces are symmetry-protected, whereas the electron-like FS remain nodeless.
    (d) Nodal $d_{xy}$ symmetry. The nodes on both the hole-like and electron-like FS are symmetry-protected.
    The sketches are based on Ref.\,\onlinecite{Hirschfeld2011RoPP}.
    }
    \label{fig:gap_symmetries}
\end{figure*}

In Ref.\,\onlinecite{Maiti2011PRL}, Maiti \textit{et al}. used a model for iron-based superconductors with three hole-like FS sheets around $\Gamma$ and two electron-like pockets around M. They approximated all the intraband and interband pairing interaction components with their leading angular harmonics in $s$-wave and $d_{x^2 -y^2}$-wave and analyzed the leading symmetry as a function of either hole or electron doping. They found that the $s_\pm$ symmetry, with isotropic gaps of different sign on the hole-like and electron-like FSs (Figure\,\ref{fig:gap_symmetries}a,b), is favoured when both electron and hole FS sheets are present, while at extreme hole or electron doping where one of the two kinds of bands disappears, the $d$-wave symmetry prevails (Figure\,\ref{fig:gap_symmetries}c,d). This is what happens at either end of the doping series in 122 iron-based compounds (KFe$_2$As$_2$ and KFe$_2$Se$_2$).  
The existence of node lines in the gap is generally revealed by the persistence of low-energy quasiparticles that, in turn, give rise to a linear low-temperature behaviour in some experimentally-accessible quantities like the magnetic penetration depth, the specific heat, the thermal conductivity, and so on. Various measurements have reported KFe$_2$As$_2$ to feature a nodal gap\,\cite{Hashimoto2010PRBb}, and the nodal symmetry seems to be inherited also by the sister compounds CsFe$_2$As$_2$\,\cite{Hong2013PRB} and RbFe$_2$As$_2$\,\cite{Zhang2015PRB}. The fact that these give rise to the building blocks of the corresponding 12442 compounds is intriguing and has stimulated the interest in the determination of the gap symmetry in this family of iron-based superconductors. As a matter of fact, stoichiometric 12442 are actually on the verge of the disappearance of the electron-like pockets due to the intrinsic hole doping. Hence, it immediately became clear that an experimental determination of the gap symmetry in these compounds could provide an important test for theoretical predictions.

As pointed out in Ref.\,\onlinecite{Hirschfeld2011RoPP}, one first test of the symmetry is provided by the spin resonance. The scattering between hole-like and electron-like FS sheets produces a resonance peak at or near $Q = (\pi, \pi)$, as experimentally observed. This peak is sharper in the case of the $s_\pm$ symmetry and more rounded for the $d$-wave symmetry. The results of inelastic neutron scattering in K-12442\,\cite{Hong2020PRL} thus support an $s_\pm$ symmetry. In principle, this does not prevent the possible presence of accidental nodes, that indeed can appear on either the electron-like or the hole-like FS sheets, in suitable conditions.  

\begin{figure*}
    \centering
    \includegraphics[width=\linewidth]{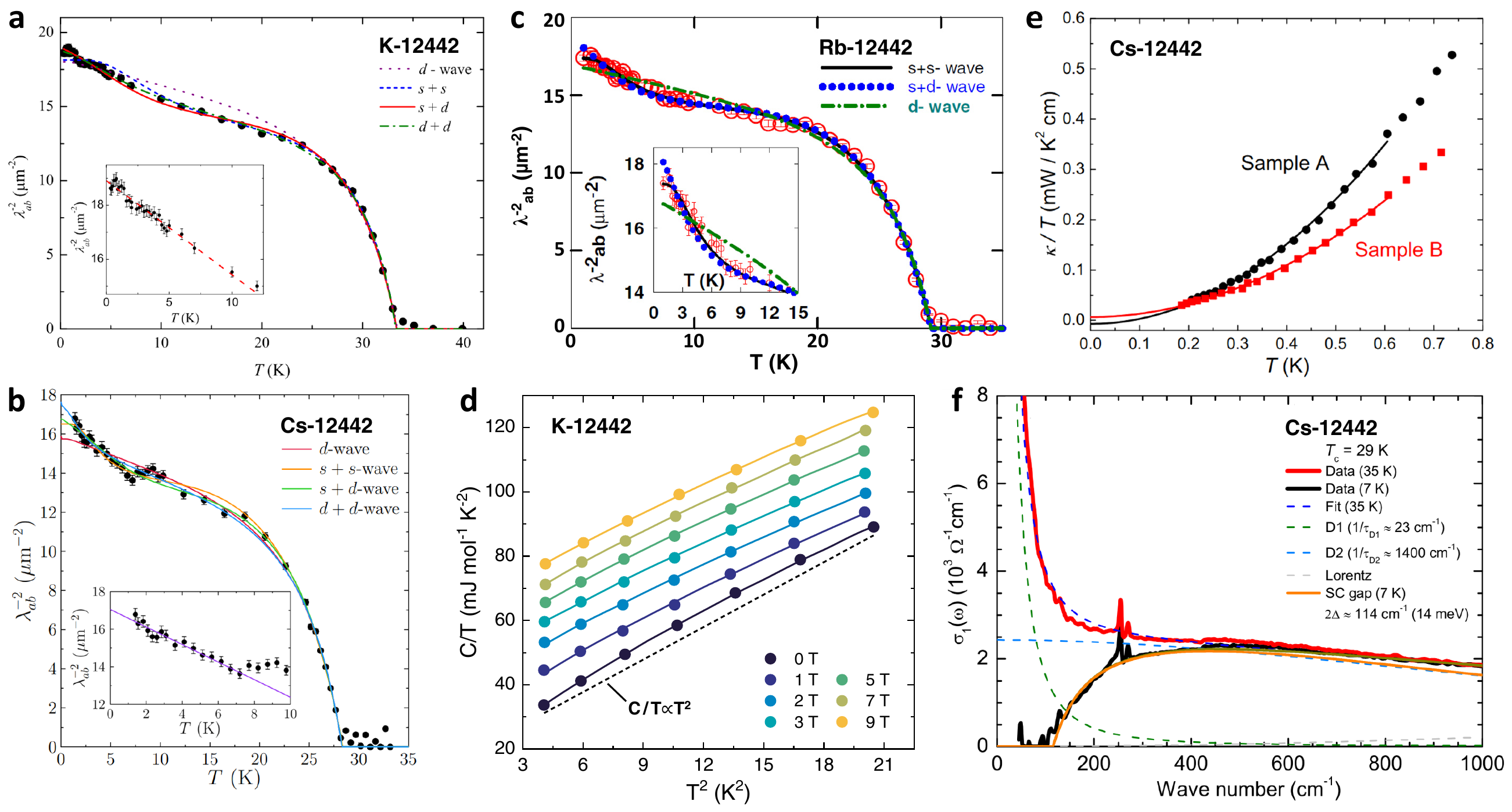}
    \caption{
    (a-c) Temperature dependence of the inverse squared penetration depth (proportional to the superfluid density) measured via muon spin rotation in K-12442\,\cite{Smidman2018PRB} (a), Cs-12442\,\cite{Kirschner2018prb} (b), and Rb-12442\,\cite{Adroja2018JPSJ} (c). Solid lines are fits to different models for the SC gap symmetry (Eq.\,\ref{eq:uSR}) as discussed in the main text. Insets show the low-temperature data in an expanded scale, highlighting the linear dependence of a gap with line nodes. Reprinted with permission from Refs.\,\onlinecite{Smidman2018PRB, Kirschner2018prb, Adroja2018JPSJ} respectively. Panels (a) and (b) are under copyright 2018 by the American Physical Society. Panel (c) is under copyright 2020 by the Physical Society of Japan.
    (d) Specific heat of K-12442 as a function of squared temperature under the application of different magnetic fields. Solid lines are fits to Eq.\,\ref{eq:SH} which highlight the presence of a quadratic term in the curves, typical of line nodes in the SC energy gap(s).
    Adapted from Ref.\,\onlinecite{Wang2020SciChi}.
    (e) Temperature dependence of thermal conductivity for two Cs-12442 single crystals in zero magnetic field. Solid curves are fits to $\kappa/T = a + bT^2$ for $T\leq0.6$\,K, showing negligible residual linear terms associated with nodal behavior. Reprinted with permission from Ref.\,\onlinecite{Huang2019PRB}. Copyright 2019 by the American Physical Society.
    (f) Optical conductivity of Cs-12442 above (35\,K, red line) and below (7\,K, thick black line) $T_c$. Dashed lines are the Drude-Lorentz fits at 35\,K. The orange solid line is the dirty-limit Mattis-Bardeen contribution with a SC energy gap of 7\,meV at 7\,K. Reprinted with permission from Ref.\,\onlinecite{Xu2019PRB}. Copyright 2019 by the American Physical Society.
    }
    \label{fig:uSR_thermal_optical}
\end{figure*}

Indeed, several experiments carried out in 12442 compounds gave evidence of nodes in the gap. Transverse-field muon spin rotation ($\upmu$SR) experiments in K-12442\,\cite{Smidman2018PRB}, Cs-12442\,\cite{Kirschner2018prb} and Rb-12442\,\cite{Adroja2018JPSJ} polycrystals (shown in Figure\,\ref{fig:uSR_thermal_optical}a, b and c respectively) clearly highlighted that the temperature dependency of $\lambda_{ab}^{-2}$, which is actually proportional to the superfluid density $\rho_S = \left[\lambda(T=0)/\lambda(T)\right]^2$, cannot be reconciled with either a single or multiple isotropic gaps. Indeed, at low temperature the superfluid density does not saturate (as one would expect in the case of a full gap) but displays a linear increase on decreasing $T$, which is expected in the case of a nodal gap. However, a clear change in slope between 5 and 10\,K cannot be reproduced within a single $d$-wave gap picture. The experimental curves can rather be fitted by an expression
\begin{equation}
    \rho_S (T) = x \rho_{S,1}^{s,d}(T)+ (1-x)\rho_{S,2}^{s,d}(T).
    \label{eq:uSR}
\end{equation}
in which the superfluid density is simply the weighted sum of two contributions, associated with different gaps. The combination of one isotropic ($s$-wave) and a nodal ($d$-wave) gap, referred to as $s+d$ in Refs.\,\onlinecite{Smidman2018PRB, Kirschner2018prb, Adroja2018JPSJ}, was found to provide the best fit. In particular, a smaller $d$-wave gap (of amplitude ranging from 1.8\,meV in K-12442 to 0.9\,meV in Rb-12442)  and a larger $s$-wave gap (of amplitude between 10 and 7.5\,meV) were obtained. 

Actually, a model with two nodal gaps (referred to as $d+d$) provides almost equal fit to the experimental curves of superfluid density. The so-called $d+d$ fit of the superfluid density curves gave gap amplitudes $\Delta\ped{d1}= 1.8$\,meV and $\Delta\ped{d2} =14$\,meV in K-12442\,\cite{Smidman2018PRB} and $\Delta\ped{d1}= 1.3$\,meV and $\Delta\ped{d2} =14$\,meV in Rb-12442\,\cite{Adroja2018JPSJ}. With a critical temperature of the order of 30\,K, a gap amplitude of the order of 14\,meV is actually difficult to accept, as it would give  a gap ratio $2 \Delta /k\ped{B} T_c \simeq 11$. 

As a general comment to the fit of the superfluid density, it is worthwhile to point out that the two-gap picture is to be intended as an effective one, since only gaps that are sufficiently different in amplitude can be experimentally disentangled by any measurement that is not band-resolved (even though the compounds feature more than just two FSs).
A more subtle issue concerns the comparison between the results of these fits and the allowed gap structures in an iron-based compound with tetragonal lattice structure, explained in Ref.\,\onlinecite{Hirschfeld2011RoPP}. In the fit of the superfluid density, the name $s+d$ does not refer here to the $s+d$ symmetry of a single gap, but to two distinct gaps residing on different FSs. Therefore, the $s+d$ model corresponds to the case in which a small gap (in some of the FS sheets) is nodal, and expressed for simplicity as a $d$-wave one, and a larger one (in other FS sheets) is not. By comparison with the more correct nomenclature of Ref.\,\onlinecite{Hirschfeld2011RoPP}, this is what happens in the nodal $s\ped{\pm}$ gap structure (Figure\,\ref{fig:gap_symmetries}b), in which accidental nodes occur (in some FS), or in the $d_{x^2-y^2}$-wave structure (Figure\,\ref{fig:gap_symmetries}c), in which symmetry-protected nodes occur on the hole-like FS (but not on the electron-like ones). The $d+d$ model, instead, would correspond to the case in which both the effective gaps display lines of nodes. Taking into account the topology of the FS, this possibility could  corresponds to the $d_{xy}$-wave gap symmetry (Figure\,\ref{fig:gap_symmetries}d), in which node lines cross both the hole-like and the electron-like FSs\,\cite{Hirschfeld2011RoPP}. Again, since the model treats the gaps on the electron-like and hole-like FSs as independent entities, both of them are modelled with the simplest nodal gap, i.e. $d$-wave. 

Specific-heat (SH) measurements in K-12422 single crystals\,\cite{Wang2020SciChi} in various magnetic fields showed a clear SH jump at $T_c$ but also highlighted that the low-temperature dependence of the SH (shown in Figure\,\ref{fig:uSR_thermal_optical}d) could not be fitted by a curve of the kind $C(T,H) = \gamma (H) T + \beta T^3$, but rather required an additional term proportional to $T^2$:
\begin{equation}
    C(T,H) = \gamma(H) T + \alpha(H) T^2 + \beta T^3.
    \label{eq:SH}
\end{equation}
The presence of the quadratic term in the electronic SH is the hallmark of line nodes in the energy gap(s) and its effect is particularly clear at lower temperatures. At zero field, a residual term $\gamma_0 = \gamma (0) \simeq 5.3 \, \mathrm{mJ\, mol^{-1} K^{-2}}$ was observed, which signals the persistence of unpaired quasiparticles at low energy. Since the superconducting fraction of the crystals was close to 100\%, this, in turn, can be associated to the existence of line nodes in the energy gap. The magnetic-field dependence of $\Delta \gamma (H) = \gamma(H)- \gamma_0$ reflects the multigap nature of the compound and suggests the existence of a large anisotropy on an individual FS sheet as well as the presence of a larger full gap on other FSs. 

Ultralow-temperature thermal conductivity measurements in Cs-12442 single crystals\,\cite{Huang2019PRB} showed a negligible residual linear term in $\kappa_0/T$ in zero field, which is generally the indication of a fully-gapped superconductor (see Figure\,\ref{fig:uSR_thermal_optical}e). The magnetic-field dependence of $\kappa_0/T$ was not compatible with a single gap and rather suggested multiple nodeless gaps (with a ratio between the large and the small one of the order of 2), being very similar to that of moderately-doped 122 compounds.

Optical spectroscopy measurements in Cs-12442 single crystals\,\cite{Xu2019PRB} were interpreted in a single-isotropic gap scenario and the extracted gap was about 7\,meV (see Figure\,\ref{fig:uSR_thermal_optical}f). 

\begin{figure}
    \centering
    \includegraphics[width=\linewidth]{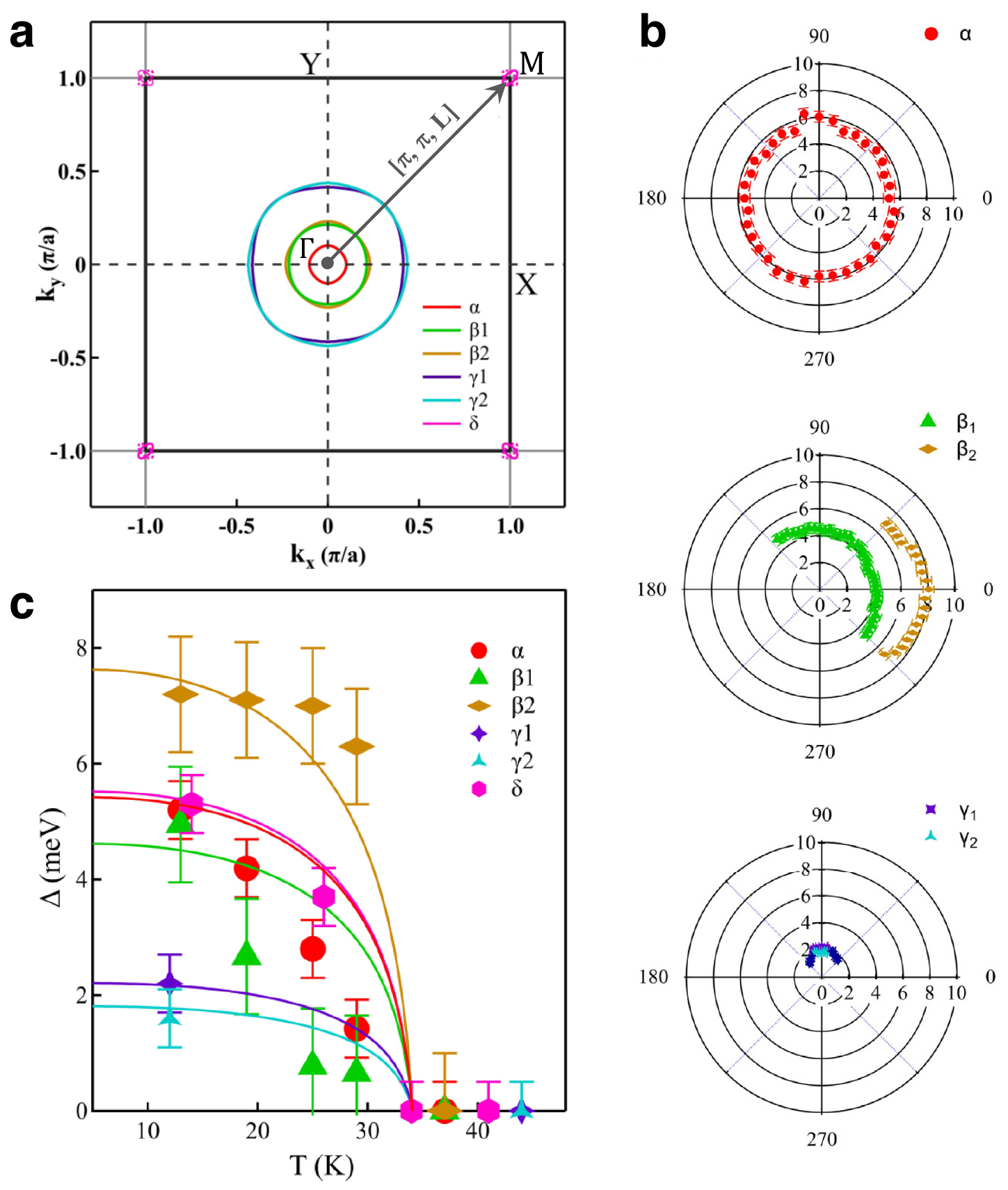}
    \caption{
    (a) Fermi surface sheets observed via angle-resolved photoemission spectroscopy (ARPES) measurements in K-12442. The spin resonance vector is depicted as a black arrow.
    (b) Angular and (c) temperature dependencies of the band-resolved SC energy gaps measured on the different hole-like Fermi sheets via ARPES. Filled symbols in (b,c) are the experimental data, solid lines in (c) are fits to the expected BCS energy dependence.
    Reprinted with permission from Ref.\,\onlinecite{Wu2020PRB}. Copyright 2020 by the American Physical Society.
    }
    \label{fig:ARPES}
\end{figure}

A direct experimental determination of the energy gaps, together with their association with the different FS sheets, was provided by high-resolution laser-based angle-resolved photoemission spectroscopy (ARPES) measurements carried out in K-12442 single crystals\,\cite{Wu2020PRB}. As already mentioned, ARPES gave evidence of several FSs (see Figure\,\ref{fig:ARPES}a): three hole-like cylinders centered about the $\Gamma$ point, of which two are nearly degenerate ($\alpha$, $\beta_1$ and $\beta_2$, $\gamma_1$ and $\gamma_2$), and tiny electron-like pockets at the M point ($\delta$). The splitting of the $\beta$ and $\gamma$ bands can be attributed to the interlayer interaction within a bilayer, as it happens in bilayer cuprates. 
The energy gaps were measured on all the FSs. Figure \ref{fig:ARPES}b shows the amplitude of the gaps on the hole-like FSs, as reconstructed from different cuts at different angles with respect to the $\Gamma-X$ direction. None of the gaps shows nodes, but they all display an in-plane anisotropy, their amplitude being systematically minimum at $\theta = 45^{\circ}$ that corresponds to the direction connecting the $\Gamma$ point with the M point (i.e. the direction of the wavevector at which the spin resonance is observed). 
  
The gap amplitude on the electron-like pockets, $\delta$, turns out to be similar to that of the $\alpha$ sheet but it could not be measured as a function of the angle due to the smallness of the relevant FS. Hence, there are actually no indications about the isotropy/anisotropy of this gap. 

The topology of the FS makes the possibility of a nesting-driven pairing mechanism very unlikely. As a consequence, the authors compared their results to the predictions of the ``strong coupling" approach, based on local (short-range) interactions. They proposed a generalized $s$-wave gap function
\begin{equation}
    \Delta_s = \left| \frac{1}{2}\Delta_0 (\cos k_x + \cos k_y) \pm \frac{1}{2} \Delta_z \cos\left( \frac{k_y}{2}\right)\right|
\end{equation}
in the 2-Fe unit cell, which was found to work in the case of (Ba$_{0.6}$K$_{0.4}$)Fe$_2$As$_2$\,\cite{Xu2011NatPhys}, and takes into account the lattice symmetry and the absence of dispersion along $k_z$. Here $\Delta_0$ originates from the intralayer next-nearest-neighbour exchange coupling $J_2$, and $\Delta_z$ from the interlayer exchange coupling $J_z$. This last term has to be taken into account in order to explain the difference in the gap amplitude between the two $\beta$ split bands. 
They observed, however, that the measured gap amplitudes fit with this function only if the parameters $\Delta_0$ and $\Delta_z$ are allowed to be band{-}dependent.
Figure\,\ref{fig:ARPES}c finally shows the temperature dependence of the energy gaps (symbols) compared to BCS-like behaviours (solid lines). 

\begin{figure}
    \centering
    \includegraphics[width=0.95\linewidth]{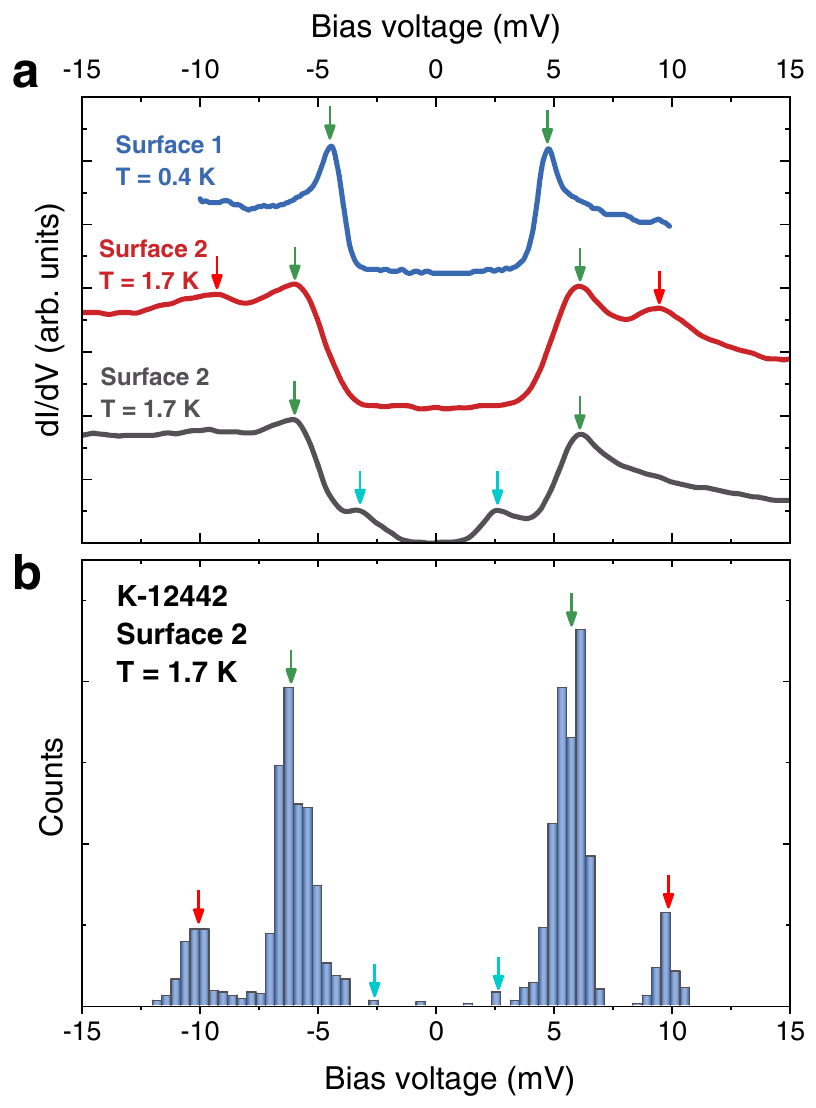}
    \caption{
    (a) Differential conductance spectra measured via scanning tunnelling spectroscopy (STS) on two different terminated surfaces of K-12442.
    (b) Statistics of the peak energies derived from 900 STS spectra measured in a grid of $30\times30$ uniformly-distributed points in a $100\times100$\,nm\apex{2} area on the second type of surface.
    In both panels, the green arrows at 4--7\,meV highlight the coherence peaks of the SC energy gap. The blue arrows at about 2.5\,meV highlight spectral features attributed to impurity bound states. The red arrows at about 10\,meV highlight humps attributed either to a larger gap or to bosonic modes.
    Adapted from Ref.\,\onlinecite{Duan2021PRB}.}
    \label{fig:STS}
\end{figure}

Recent measurements of the energy gap in K-12442 single crystals by means of tunnelling spectroscopy\,\cite{Duan2021PRB} gave results that are summarized in Figure\,\ref{fig:STS}. The authors of Ref.\,\onlinecite{Duan2021PRB} claimed that, after cleaving the crystal in vacuum, two kinds of terminated surfaces were exposed. In the most common type of surface, STS measurements clearly highlighted a single full gap, uniform in space, of about 4.6\,meV, and (in some spectra) additional features at about 2.2\,meV that the authors associated to impurity-induced bound-state peaks. On the second type of surface, instead, multiple gaps could be obtained from either single spectra or separate ones. The distribution of gap amplitudes ranges from 4 to 7\,meV, and additional features are observed at higher energies (around 10\,meV). Figure\,\ref{fig:STS}a reports some of the spectra measured on either kind of surface, and Figure\,\ref{fig:STS}b the distribution of the gap values found in the second kind of surface. It is not clear why STS does not detect the gap of about 2\,meV that ARPES associates to the largest hole-like Fermi cylinder.

\section{Effect of substitutions in the F\lowercase{e} site}

\begin{figure}
    \centering
    \includegraphics[width=\linewidth]{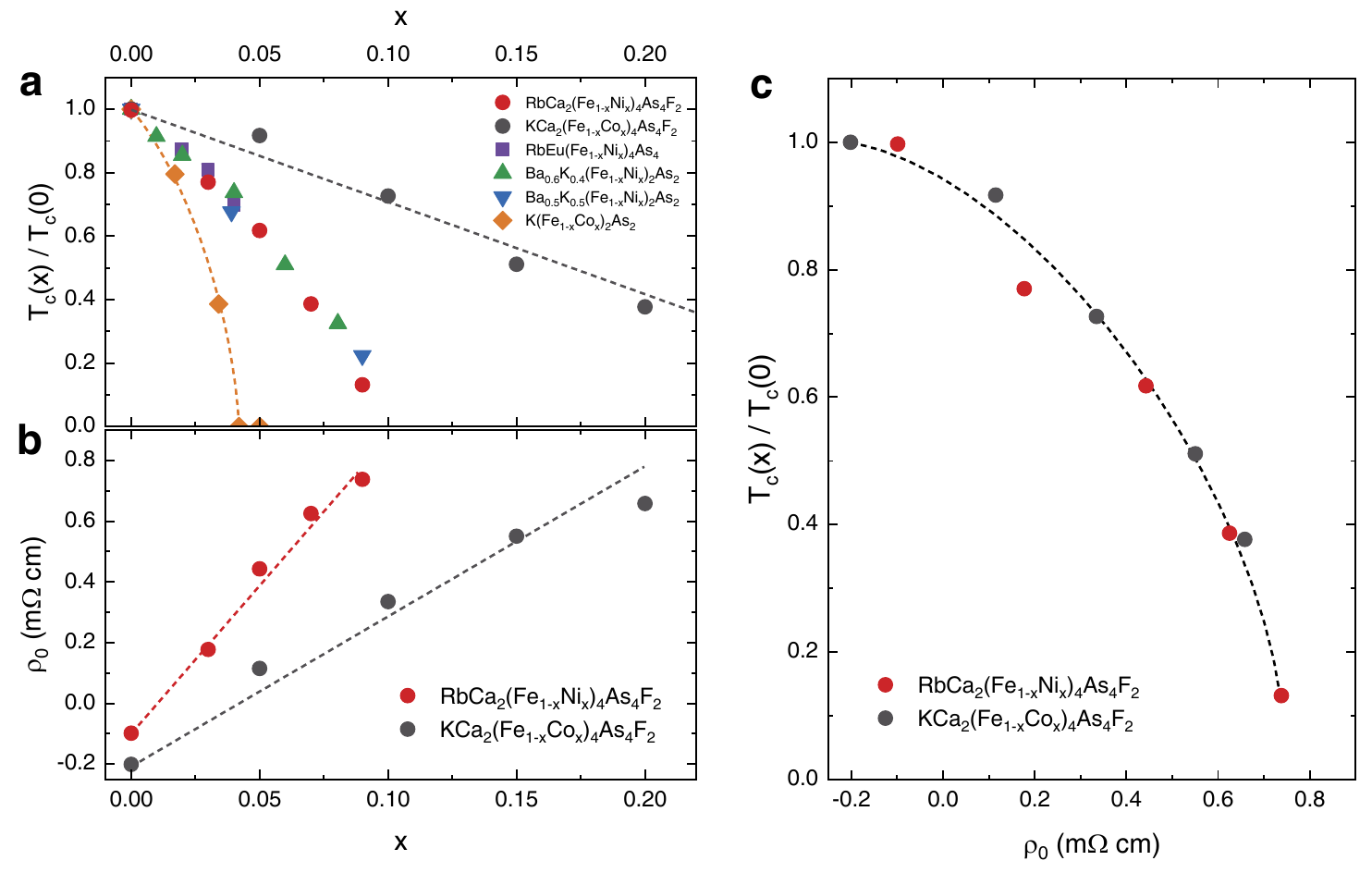}
    \caption{
    (a) Normalized SC transition temperature as a function of substitutional doping at the Fe site for Ni-doped Rb-12442\,\cite{Yi2020NJP}, Co-doped K-12442\,\cite{Ishida2017PRB}, Ni-doped RbEu-1144\,\cite{Willa2020PRB} and BaK-122\,\cite{Cheng2013EPL, Li2012PRB}, and Co-doped nodal compound K-122\,\cite{Wang2014PRB}.
    (b) Residual resistivity $\rho_0$ of Ni-doped Rb-12442\,\cite{Yi2020NJP} and Co-doped K-12442\,\cite{Ishida2017PRB} as a function of substitutional doping, obtained by fitting the low-temperature resistivity data to a power law $\rho(T)=\rho_0+AT^n$.
    (c) Normalized SC transition temperature as a function of the residual resistivity in Ni-doped Rb-12442 and Co-doped K-12442.
    Panels (a) and (b) are adapted from Ref.\,\onlinecite{Yi2020NJP}. Dashed lines are guides to the eye.}
    \label{fig:Fe_doping}
\end{figure}

Another important piece of information about the 12442 compounds comes from the study of the effect of substitutions in the Fe site. In particular, Ref.\,\onlinecite{Ishida2017PRB} and Ref.\,\onlinecite{Yi2020NJP} reported systematic studies of the effect of Ni substitution in Rb-12442 and Co substitution in K-12442.
In either case, the substitution results in an effective electron doping and, as evidenced by the sign change in the Hall coefficient $R\ped{H}$, makes the majority carriers change from holes to electrons. At $x=0.06$ (for Ni-doped Rb-12442) and $x=0.1$ (for Co-doped K-12442), $R\ped{H}$ is zero and this implies that, at this doping content, the compound becomes compensated. As evidenced by density-functional theory calculations of the FS\,\cite{Ishida2017PRB}, these substitutions also produce a shrinking of the hole cylinders about the $\Gamma$ point, accompanied by an expansion of the electron pockets. Near compensation, this produces almost perfect nesting conditions between hole-like and electron-like bands, a condition that, however, does not correspond to the onset of a SDW order -- reasonably due to the intrinsic disorder\,\cite{Ishida2017PRB}. 
Moreover, a comparison between RbCa$_2$(Fe$_{1-x}$Ni$_x$)$_4$As$_4$F$_2$ and KCa$_2$(Fe$_{1-x}$Co$_x$)$_4$As$_4$F$_2$ shows that, for any given concentration $x$, Ni substitution in Rb-12442 results in a larger suppression of the critical temperature (with respect to the undoped compound) than Co doping in K-12442 (see Figure\,\ref{fig:Fe_doping}a)
. The $T_c$ vs $x$ curves are indeed rather different\,\cite{Yi2020NJP}; the one pertaining to Co-doped K-12442 has a rather small (negative) slope, while the one measured in Ni-doped Rb-12442 is steeper, and very similar to that obtained in iron-based compounds with a similar initial hole doping content, belonging to the 1144 family (RbEu(Fe$_{1-x}$Ni$_x$)$_4$As$_4$\,\cite{Willa2020PRB}) and to the 122 family (i.e. Ba$_{0.5}$K$_{0.4}$(Fe$_{1-x}$Ni$_x$)$_2$As$_2$\,\cite{Cheng2013EPL}  and Ba$_{0.5}$K$_{0.5}$(Fe$_{1-x}$Ni$_x$)$_2$As$_2$\,\cite{Li2012PRB}). The $T_c$ vs $x$ curve of the nodal compound K(Fe$_{1-x}$Co$_x$)$_2$As$_2$ is instead much steeper\,\cite{Wang2014PRB}. Similarly, also the residual resistivity $\rho_0$ (here obtained from the fit of the low-temperature resistivity data with a function $\rho (T) = \rho_0 + A T^n$) increases much more upon Ni doping than upon Co doping, as shown in Figure\,\ref{fig:Fe_doping}b.

It is widely known that the effect of disorder on the critical temperature can provide some hints about its gap symmetry. In particular, in multiband systems, interband scattering from non-magnetic impurities is expected to dramatically suppress superconductivity when the order parameter changes sign between different bands. Even though the situation is actually complicated\,\cite{Hirschfeld2011RoPP}, a basic point is that the dopant concentration is not necessarily a good indicator of the level of disorder. Indeed, if one plots the $T_c$ of RbCa$_2$(Fe$_{1-x}$Ni$_x$)$_4$As$_4$F$_2$ and KCa$_2$(Fe$_{1-x}$Co$_x$)$_4$As$_4$F$_2$ as a function of the residual resistivity $\rho_0$ (see Figure\,\ref{fig:Fe_doping}c), the two curves are exactly superimposed. This means that there is actually no difference between these two compounds as far as the suppression of $T_c$ by substitutional disorder is concerned.  

\section{Point-contact measurements in R\lowercase{b}-12442}
Most of the experimental investigations of the gap symmetry carried out so far refer to the K-based 12442 compound, that features the highest $T_c$, and a few of them to the Cs-based one. The Rb-based 12442 compound is by far the least studied of the series. Since the K-Rb-Cs 12442 series does show variations in some lattice parameters, we think it is not completely sure whether the tiny details of the gap structure are preserved in all the three compounds. In particular, the fact that it was not possible to assess the angle-dependence of the gap amplitude on the electron pocket leaves the possibility for nodes to show up on that FS even if the overall results of ARPES measurements are true for all the compounds of the family. More dramatic differences with respect to the case of K-12442 could be driven by a possible further shrinkage of the electron-like pocket or by its disappearance. In that case, a transition to a $d$-wave gap would be expected, even though first-principles calculations of the bandstructure\,\cite{Ghosh2020} predict no significant difference between K, Rb and Cs-based 12442 compounds.
Motivated by these arguments, we decided to study the gap symmetry in high-quality single crystals of RbCa$_2$Fe$_4$As$_4$F$_2$. We furthermore also measured Ni-doped crystals in order to study the effect of disorder and electron doping on the order parameter.

\subsection{Experimental}

\begin{figure}
    \centering
    \includegraphics[width=\linewidth]{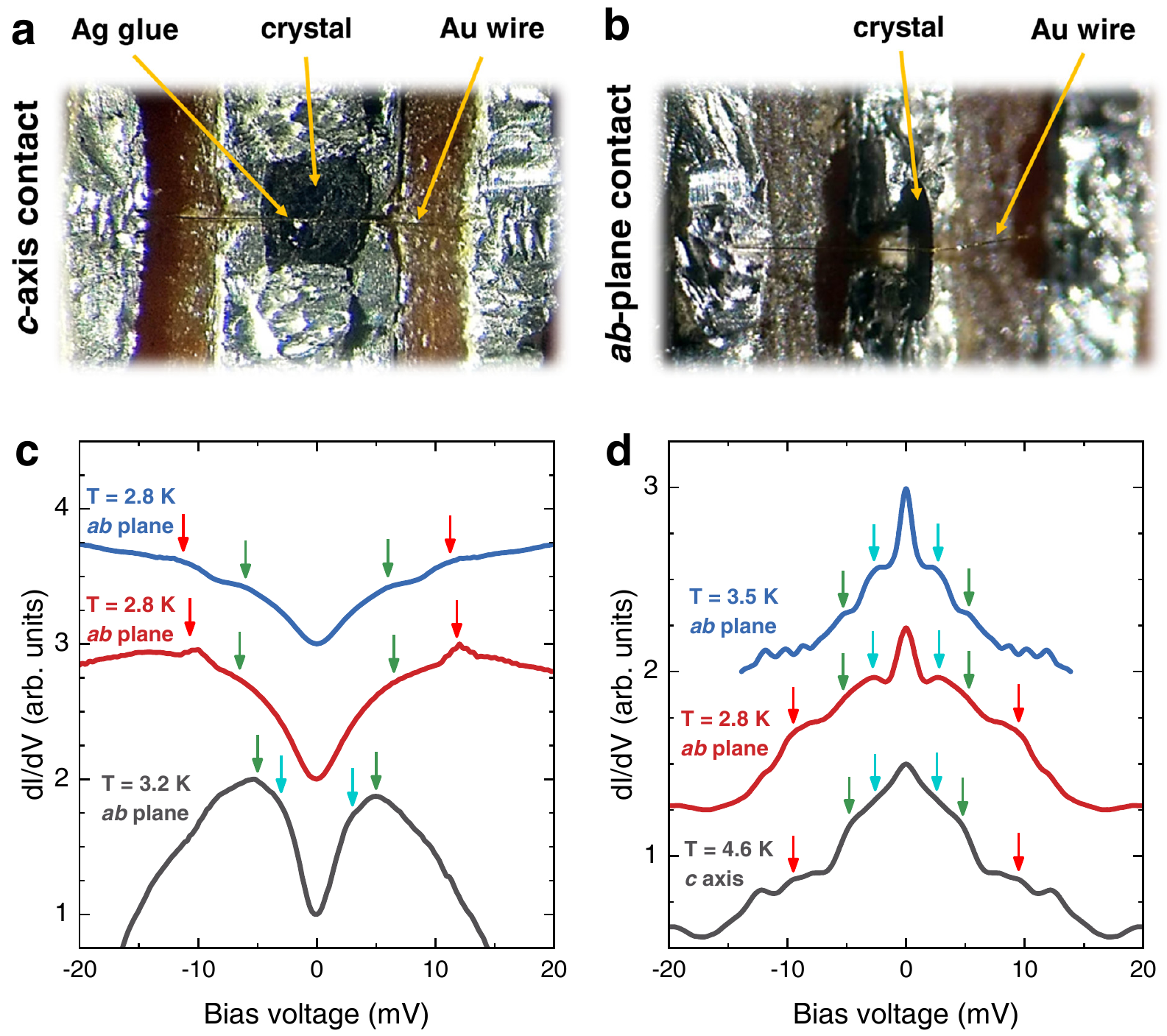}
    \caption{
    (a) Photograph of a $c$-axis soft point contact realized with Ag paste in a flat, mirror-like area of a Rb-12442 single crystal.
    (b) Photograph of an $ab$-plane soft point contact realized without Ag paste along the sharp edge of the same crystal.
    (c) Three representative examples of point-contact differential conductance spectra acquired along the $ab$ plane of Rb-12442 crystals in the tunnelling regime. All spectra show a V-shaped minimum.
    (d) Three representative examples of point-contact differential conductance spectra acquired either along the $ab$ plane or the $c$ axis of Rb-12442 crystals in the Andreev-reflection regime. All spectra show either a peak ($ab$ plane) or a cusp ($c$ axis) at zero bias, indicating the presence of nodes on at least one energy gap.
    In both (c,d), the blue, green and red arrows highlight the spectral features associated to the small and large SC energy gap, and the high-energy structures possibly associated to bosonic modes respectively. {\color{blue}All the spectra have been vertically rescaled and shifted  to facilitate the visual comparison of the position of the structures; therefore, the units on the vertical axes are arbitrary.}
    Some of the data is adapted with permission from Ref.\,\onlinecite{Torsello2022npjQM}.}
    \label{fig:PCARS_method}
\end{figure}

The single crystals of RbCa$_2$(Fe$_{1-x}$Ni$_x$)$_4$As$_4$F$_2$ were synthesized by the self-flux method using RbAs\,\cite{Wang2019JPCC, Xing2020sust, Yi2020NJP}. A detailed study of the structural and chemical aspects of the crystals can be found in Ref.\,\onlinecite{Yi2020NJP} and in Ref.\,\onlinecite{Torsello2022npjQM}. For the following of the discussion, the most important feature of these crystals is their phase purity, that allows us to exclude the possibility of contributions from residual nodal RbFe$_2$As$_2$ phases. 

The crystals are very thin platelets (whose thickness is often of the order of some tens of microns) with shiny surfaces. Figure\,\ref{fig:PCARS_method}a,b shows the mounting of a crystal for point-contact measurements along the $c$ axis (panel a) and along the $ab$ planes (panel b). Here, the name indicates the direction of the normal to the N/S interface or, more understandably although less correctly, the direction of main current injection into the sample. In either case, the crystal was stuck between two indium blocks that allow the electrical contact and also ensure mechanical support. The point contact was made by using a thin Au wire ($\varnothing = 12\, \upmu$m) stretched over the sample. In some cases (see for example Figure\,\ref{fig:PCARS_method}a), we used a tiny drop of Ag conducting paste to keep the wire in position and ensure an electrical contact with the sample surface. In other cases, the contact was directly made between the Au wire and the sample (as in Figure\,\ref{fig:PCARS_method}b). In either case, several parallel nano-contacts are very likely to be established between the crystal and the normal electrode\,\cite{Daghero2011}.  
The contact configuration is the pseudo-four-probe one, with $I^+$ and $V^+$ contacts connected to the Au wire, and the $I^-$ and $V^-$ contacts connected to either In stack.

Performing directional point-contact measurements in both the $ab$-plane and the $c$-axis configurations is very useful (if not mandatory) whenever the $\mathbf{k}$ dependence of the order parameter is suspected not to be trivial\,\cite{Daghero2013LTP}. Avoiding unwanted contributions due to the conduction along crystallographic directions that are different from the ``nominal" one is thus important. For this reason, $c$-axis contacts were made on the flatter, mirror-like parts of the topmost surface, avoiding terrace edges in order to minimize the contribution of $ab$-plane conduction. For the $ab$-plane contacts, we exploited the fact that our crystals showed regular side surfaces, flat and with sharp edges. This ensures that the direction of current injection is fairly well controlled, in the sense that it mainly occurs along the $ab$ plane, even though some contribution from the $c$ axis is possible. Instead, we cannot say what is the orientation of current injection \emph{within} the $ab$ plane. 

Figure\,\ref{fig:PCARS_method}c shows some spectra on undoped crystals, measured in $ab$-plane contacts, and displaying a predominant tunnel character. The spectra display a $V$-shaped minimum at zero bias, and clear changes of slope at two different energies, i.e $5\div 6$\,meV (green arrows) and $2.7 \div 3.0$\,meV (cyan arrows) that can be interpreted as the hallmarks of two gaps of different amplitude. These values compare well with the positions of the maxima in STS curves measured in K-12442 shown in Figure\,\ref{fig:STS}a. Exactly as in the STS spectra, additional features at about 10\,meV can be clearly seen (red arrows). 
The spectra shown in Figure\,\ref{fig:PCARS_method}d were instead measured either along the $ab$ plane or the $c$ axis (as indicated by labels) and clearly display an enhancement of conductance at low energies that can be attributed to Andreev reflection. Clearly, all these spectra (as 100\% of the Andreev spectra in Rb-12442) present a conductance maximum at zero bias, and clear shoulders at $\pm 2.5$\,meV (cyan arrows),  $\pm 5$\,meV (green arrows) and $\pm 10$\,meV (red arrows), thus compatible with the observed structures in the tunneling regime.  The persistence of these structures at the same energy in all the contacts suggests that the zero-bias maximum is not due to heating effects (since in that case, no spectroscopic features would be observed) and that the local minima are not  ``dips'' due to the critical current \cite{Sheet2004PRB,Doring2014JPCS} (since their position would depend con the contact resistance). Rather, it is strongly suggestive of a gap with nodes. Its physical origin is however different depending on the direction of the current injection.



\subsection{Origin of the zero bias maximum in \emph{ab}-plane and \emph{c}-axis contacts}

\begin{figure}
    \centering
    \includegraphics[width=\linewidth]{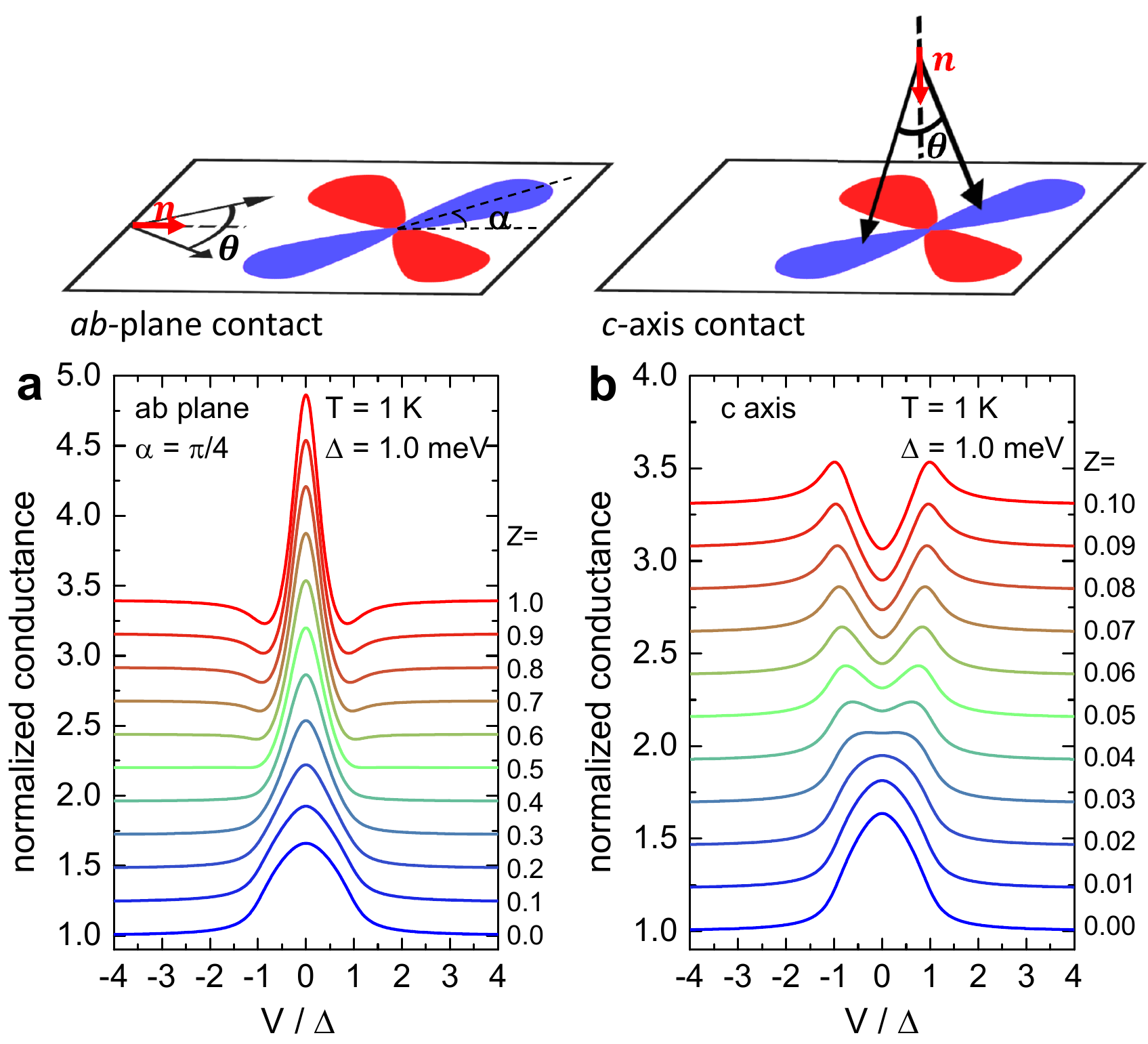}
    \caption{ 
    (a) Theoretical differential conductance spectra calculated at $T=1$\,K by using the model of Ref.\,\onlinecite{KashiwayaPRB1996} for a $d$-wave gap, in the case of in-plane injection along the nodal direction ($\alpha = \uppi/4$). The parameter $Z$ expresses the height of the potential barrier at the interface, i.e. $Z=0$ corresponds to a direct N/S junction, while $Z\to \infty$ corresponds to an ideal S/I/N junction. On increasing $Z$, the zero-bias maximum due to the zero-energy quasiparticles evolves toward a zero-bias peak due to Andreev bound states.
    (b) Same as in (a), but for $c$-axis current injection. Here, the model used is that reported in Ref.\,\onlinecite{YamashiroPRB1997}, in which an integration is made over a thin cylindrical belt which represents the 2D FS. The values of $Z$ are much smaller than in the other case because the small  angular region of integration enhances the effect of the barrier parameter. No peaks associated to the Andreev bound states are predicted for any $Z$ because ELQ and HLQ always ``feel'' the same order parameter.
        }
    \label{fig:PCARS_theor}
\end{figure}

The model for Andreev reflection at the interface between a normal metal and an anisotropic superconductor was first proposed by S. Kashiwaya and Y. Tanaka \cite{TanakaPRL1995,KashiwayaPRB1995,KashiwayaPRB1996} as an extension of a previous model by Blonder, Tinkham and Klapwijk (BTK model)\,\cite{BlonderPRB1982}. This model is essentially 2D and predicts that, when the injection occurs in the basal plane, an electron  incident on the N/I/S interface with a wavevector $\mathbf{k}_\mathrm{N}$ that makes an angle $\theta$ with the normal to the interface, $\mathbf{n}$, can undergo four processes, whose probabilities depend on the height of the potential barrier at the interface (modelled by the dimensionless parameter $Z$), and on the energy of the electron: a) Andreev reflection, that consists in the transmission of a Cooper pair in S and in the retroreflection of a hole in N; b) normal reflection in N; c) normal transmission as an electron-like quasiparticle (ELQ) at an angle $\theta$ with $\mathbf{n}$; d) transmission as a hole-like quasiparticle (HLQ) at an angle $\uppi-\theta$ with $\mathbf{n}$. 
A schematic picture of these processes is given for example in Ref.\,\onlinecite{KashiwayaPRB1996}. 

If the superconducting order parameter depends on $\mathbf{k}$, HLQ and ELQ may feel different values of it, $\Delta_\mathrm{H}$ and $\Delta_\mathrm{E}$. For example, let us assume that the gap has a $d$ symmetry; if $\mathbf{n}$ is parallel to the antinodal direction, HLQ and ELQ feel the same gap irrespective of  the angle of incidence $\theta$. But, if $\mathbf{n}$ makes an angle $\alpha\neq 0$ with respect to the antinodal direction, $\Delta_\mathrm{H}$ and $\Delta_\mathrm{E}$ will have different sign for some values of $\theta$. This gives rise, for any $Z>0$, to zero-energy states at the surface (called Andreev bound states) which arise from the positive interference between HLQ and ELQ\,\cite{KashiwayaPRB1995,TanakaPRL1995,KashiwayaPRB1996}. These states contribute to the conduction through the junction and give rise to a zero-energy peak \cite{Kashiwaya2000RoPP}. 

Figure\,\ref{fig:PCARS_theor}a shows some examples of theoretical spectra calculated in the case $\alpha=\uppi/4$, at finite temperature $T=1\,\mathrm{K}$, i.e. assuming a single $d_{x^2-y^2}$ gap of amplitude $\Delta=1\, \mathrm{meV}$, and for different values of the barrier parameter $Z$. The $Z=0$ spectrum shows a zero-bias maximum that is due to the availability of low-energy unpaired quasiparticles, since in the absence of a potential barrier at the interface the only possible processes are Andreev reflection and normal transmission as ELQ, and therefore there is no quasiparticle interference\,\cite{KashiwayaPRB1996, TanakaPRL1995, KashiwayaPRB1995}. On increasing $Z$, the zero-bias maximum progressively turns into a true peak whose height increases and whose width instead decreases. For $\alpha \simeq \uppi/8$, the zero-bias peak is accompanied by two symmetric maxima at the gap edges. Some additional examples of calculated spectra can be found in Ref.\,\onlinecite{DagheroPuCoGa} and in Ref.\,\onlinecite{Daghero2011}.

When the current is injected perpendicular to the basal plane, as in $c$-axis contacts, HLQ and ELQ (that are injected symmetrically with respect to the $k_z$ axis) always feel order parameters with the same sign, as shown in the schematic picture of Figure\,\ref{fig:PCARS_theor}b. In this case, therefore, there is \emph{no} quasiparticle interference and the zero-bias peak never occurs\,\cite{TanakaPRL1995, KashiwayaPRB1995}. Owing to the cylindrical FSs observed by ARPES in Rb-12442, to model this situation we used the approach described in Ref.\,\onlinecite{YamashiroPRB1997} in which the integration is made over a thin cylindrical belt, symmetric about the basal plane (i.e. over all the azimuthal angles $\phi$ and a small interval of inclination angles $\theta$ centered about $\uppi/2$). The limitation in the angles of integration, dictated by the geometry of the system, enhances the effect of the $Z$ parameter very much, so that the values of $Z$ that are usually required to fit our $c$-axis spectra are much smaller than those usually required to fit the $ab$-plane ones. This is simply a consequence of the different integration domains and does not mean that the potential barrier is different along the two directions. Figure\,\ref{fig:PCARS_theor}b reports some theoretical spectra, calculated for a $c$-axis contact using the same parameters as in panel a. The $Z=0$ spectrum is very similar to that obtained in the $ab$-plane case, but on increasing $Z$ a V-shaped zero-bias minimum progressively sets in.

\subsection{Fit of the point-contact spectra}
The zero-bias structures of the point-contact spectra prevent any fit with one or two nodeless gaps. Therefore, we first tried to fit them with a two-band effective model in which one gap is nodal and the other is nodeless. The nodal gap was expressed for simplicity as a $d$-wave one, and thus we will call this model $s{-}d$. 
{\color{blue}We thus used a two-band version\,\cite{DagheroSUST2010, Daghero2011} of the 2D-BTK model\,\cite{BlonderPRB1982, KashiwayaPRB1996} in which the differential conductance due to Andreev reflection is calculated separately for the two bands, and the total conductance is just a weighted sum of the two contributions: 
\begin{equation}
    G (E) = w_1 G_1 (E) + w_2 G_2 (E). \label{eq:weight}
\end{equation}
where $w_1 + w_2 = 1$. Here, $G_i(E)$ is the contribution to the normalized conductance due to the $i$-th gap. Its actual expression involves an angular integration whose form depends on whether the current is injected in the $ab$ plane or along the $c$ axis [see Equation (4) in Ref.\,\onlinecite{YamashiroPRB1997}].
Equation\,\eqref{eq:weight} is clearly an approximation, which is correct only in some conditions, as discussed in Ref.\onlinecite{Daghero2013LTP} (e.g. it was shown to hold in MgB$_2$\,\cite{Brinkman2002PRB}), because the correct expression of the total conductance (at $T=0$) is instead\,\cite{DagheroSUST2010,Daghero2013LTP} 
\begin{equation}\label{eq:G_FS}
\langle
G(E)\rangle_{I\parallel\mathrm{\mathbf{n}}}=\frac{\sum_{i}\left\langle\sigma_{i\mathrm{\mathbf{k}}n}(E) \tau_{\mathit{i}\mathrm{\mathbf{k}},n} \frac{v_{\mathit{i}\mathrm{\mathbf{k}},n}}{v_{\mathit{i}\mathrm{\mathbf{k}}}}\right\rangle_{\mathrm{FS}_{i}}}{\sum_{i}\left\langle\tau_{\mathit{i}\mathrm{\mathbf{k}},n} \frac{v_{\mathit{i}\mathrm{\mathbf{k}},n}}{v_{\mathit{i}\mathrm{\mathbf{k}}}}\right\rangle_{\mathrm{FS}_{i}}}
\end{equation}
Here, $\sigma_{i \mathrm{\mathbf{k}} n}(E)$ is the normalized differential conductance, calculated by using the 1D BTK model\,\cite{BlonderPRB1982}, in the $i$-th band and for the wavevector $\mathrm{\mathbf{k}}$, when the current is injected along $\mathrm{\mathbf{n}}$. The symbol $\langle \dots \rangle_{\mathrm{FS}_i}$ indicates an integral over the $i$-th sheet of the FS, and $v_{\mathit{i}\mathrm{\mathbf{k}},n}=\mathbf{v}_{\mathit{i}\mathrm{\mathbf{k}}} \cdot \mathbf{n}$ is the projection along $\mathbf{n}$ of the Fermi velocity on the $i$-th band and at wavevector $\mathrm{\mathbf{k}}$ . The normal-state barrier transparency $\tau_{\mathit{i}\mathrm{\mathbf{k}},n}$ is given by

\begin{equation}\label{eq:tau2}
\tau_{\mathit{i}\mathrm{\mathbf{k}},n}=\frac{4v_{\mathit{i}\mathrm{\mathbf{k}},n}v_{\mathrm{N},n}}{(v_{\mathit{i}\mathrm{\mathbf{k}},n}+v_{\mathrm{N},n})^{2}+4Z^{2}v_{\mathrm{N}}^{2}}
\end{equation}
where $v_{\mathrm{N},n}= \mathbf{v}_\mathrm{N} \cdot \mathbf{n}$, $\mathbf{v}_\mathrm{N}$ being the  Fermi velocity in the normal bank, and $Z$ is the barrier parameter.

The function $\sigma_{i \mathrm{\mathbf{k}} n}(E)$ that appears in Equation\,\eqref{eq:G_FS} contains the gap $\Delta_i(\mathrm{\mathbf{k}})$ and the normal-state barrier transparency $\tau_{\mathit{i}\mathrm{\mathbf{k}},n}$. Therefore, in general $\sigma_{i \mathrm{\mathbf{k}} n}(E)$ \emph{does} depend on $\mathrm{\mathbf{k}}$  and cannot be taken out of the integral at the numerator\,\cite{DagheroSUST2010,Daghero2013LTP}. On the other hand, since the material possesses several bands and we do not know exactly on which of these bands the gap(s) we measure reside, we cannot properly evaluate the Fermi velocities and their components.  Thus, we simply cannot use Equation\,\eqref{eq:G_FS} and we must make a rough, but unavoidable, approximation and use Equation\,\eqref{eq:weight} instead. The weights that appear in such equation must be taken as phenomenological parameters that certainly depend on the geometry of the FS and on the direction of current injection, but unfortunately in a way which is difficult, if not impossible, to understand in detail. The "rule of thumb" of associating the weight of a FS sheet in the total conductance to the area of its  projection on a plane perpendicular to $\mathrm{\mathbf{n}}$ is also not correct here, because it only holds in the case $Z=0$ and for isotropic gaps\,\cite{Mazin1999PRL,Daghero2013LTP}.

Using Equation\,\eqref{eq:weight} implies that the BTK fitting parameters $\Delta_i$ and $Z_i$ only appear inside the expression of $G_i$. As usual, we also introduced a broadening parameter $\Gamma_i$ for each of the gaps\,\cite{Srikanth1992, Plecenik1994} that appears as an imaginary part of the energy. This broadening parameters accounts for intrinsic effects (i.e. due to the finite quasiparticle lifetime) but also, and much more significantly, for \emph{extrinsic} effects related, for example, to inelastic quasiparticle scattering processes occurring near the N/S
interface and due, in turn, to surface degradation, contamination, etc.  In general, large $\Gamma$ values are necessary when the experimental Andreev signal is significantly reduced with respect to the theoretical prediction of the 2D-BTK model.
}


\begin{figure*}
    \centering
    \includegraphics[width=0.8\textwidth]{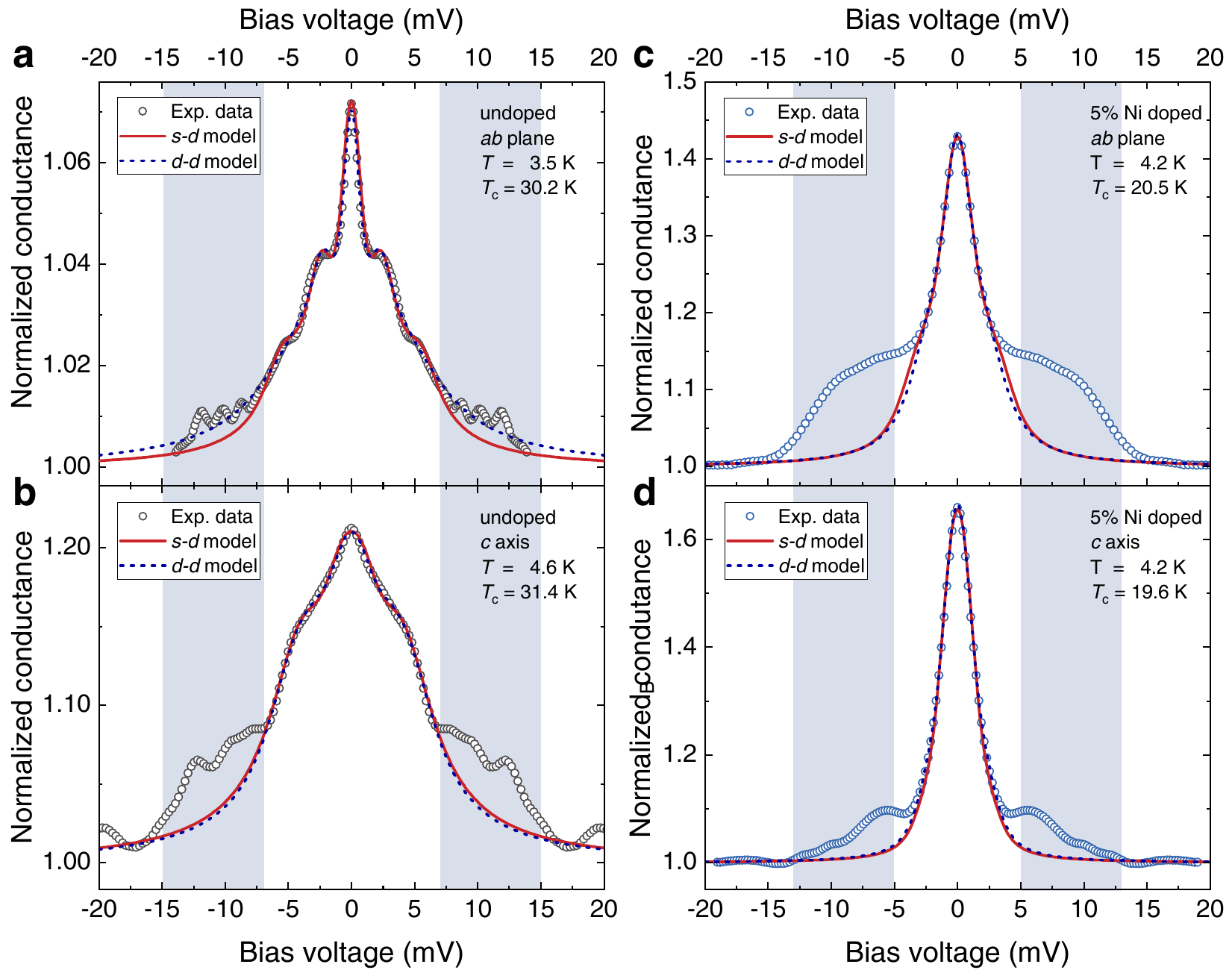}
    \caption{ 
        Low-temperature directional PCARS spectra (symbols) on single crystals of RbCa$_2$(Fe$_{1-x}$Ni$_x$)$_4$As$_4$F$_2$ with $x = 0$ (a, b) and $x = 0.05$ (c, d), and corresponding two-gap fits of the spectra either in an $s{-}d$ model (solid red lines) or in a $d{-}d$ model (dashed blue lines). All fitting parameters are summarized in Table\,\ref{tab:PCARS}.
        Adapted from Ref.\,\onlinecite{Torsello2022npjQM}.
        }
    \label{fig:PCARS_fits}
\end{figure*}

{\color{blue}Figure\,\ref{fig:PCARS_fits}a reports the normalized spectrum (symbols) of an $ab$-plane contact on undoped Rb-12442 (i.e. the compound with no Ni substitution) with the relevant $s-d$ fit (red solid curve)}. Here, the $s$-wave gap is responsible for the shoulders at about 5\,meV, while the lower-energy maxima and the zero-bias peak are both ascribed to the $d$-wave gap. This happens, indeed, when the current is injected in the $ab$ plane at an angle $\alpha < \uppi/4$ with respect to the antinodal direction (here $\alpha= \uppi/9$).  An example of $c$-axis spectrum in undoped Rb-12442 is shown in Figure\,\ref{fig:PCARS_fits}b. Here, the $s$-wave gap gives rise to the shoulders around 5\,meV and the $d$-wave gap to the zero-bias cusp, that cannot be obtained otherwise. Note that the fit unambiguously requires that the \emph{smaller gap is nodal} and the larger one is nodeless. {\color{blue}The fitting parameters corresponding to the $s-d$ fit of Figure\,\ref{fig:PCARS_fits}a and \ref{fig:PCARS_fits}b are listed in Table\,\ref{tab:PCARS}. The weight there indicated refers to the $d$-wave gap, and is higher in the $ab$-plane case than in the $c$-axis one. The broadening parameters are smaller than the corresponding gap, although their values are by far too high for the broadening to be intrinsic -- rather, they probably account for surface degradation or, even more probably, for the fact that, while we can use a model with at most \emph{two} gaps, there is a distribution of gap values\,\cite{DagheroSUST2010} with rather similar amplitude on different Fermi surfaces, as suggested by ARPES experiments\,\cite{Wu2020PRB}. } 

\begin{table}
  \begin{tabular*}{\linewidth}{cc @{\extracolsep{\fill}}cccc}
    \toprule
    & &	\multicolumn{2}{c}{Undoped} & \multicolumn{2}{c}{5\%\,Ni-doped} \\
    \midrule
    & & $ab$ plane 	& $c$ axis & $ab$ plane 	& $c$ axis	\\
    & & (Fig.\,\ref{fig:PCARS_fits}a)	& (Fig.\,\ref{fig:PCARS_fits}b) & (Fig.\,\ref{fig:PCARS_fits}c)	& (Fig.\,\ref{fig:PCARS_fits}d) \\
    \midrule
    \vspace{1mm}
    $s{-}d$ & model & & & & \\
    $\Delta_s$ & (meV) & 5.9 & 4.5 & 3.9 & 3.0\\
    $\Delta_d$ & (meV) & 3.0 & 1.9 & 1.9 & 1.45\\
    $\Gamma_s$ & (meV) & 0.7 & 1.9 & 0.7 & 0.3\\
    $\Gamma_d$ & (meV) & 0.6 & 0.5 & 0.5 & 0.075\\
    $Z_s$ &  & 0.37 & 0.32 & 0.1 & 0.1\\
    $Z_d$ &  & 0.8 & 0.5 & 0.4 & 0.002\\
    $w\ped{d}$ & & 0.9 & 0.34 & 0.65 & 0.85 \\
    $\alpha$ & & $20^\circ$ & / & $45^\circ$ & / \\
    \midrule
    \vspace{1mm}
    $d{-}d$ & model & & & & \\
    $\Delta_1$ & (meV) & 4.8 & 5.5 & 3.5 & 3.1\\
    $\Delta_2$ & (meV) & 3.0 & 1.5 & 1.5 & 1.2\\
    $\Gamma_1$ & (meV) & 3.2 & 1.6 & 1.2 & 0.4\\
    $\Gamma_2$ & (meV) & 0.6 & 0.5 & 0.1 & 0.001\\
    $Z_1$ &  & 0.95 & 0.04 & 0.1 & $\simeq 0$\\
    $Z_2$ &  & 0.8 & 0.002 & 0.47 & $\simeq 0$\\
    $w\ped{2}$ & & 0.1 & 0.2 & 0.2 & 0.5 \\
    $\alpha$ & & $20.81^\circ$ & / & $45^\circ$ & /\\
    \bottomrule
  \end{tabular*}
  \caption{
  Summary of the fitting parameters of the two-gap BTK curves shown in Figure\,\ref{fig:PCARS_fits}, for both the $s{-}d$ and the $d{-}d$ model. $\Delta_i$ are the superconducting energy gaps, $\Gamma_i$ the broadening parameters, 
  $w_i$ the weight of the small nodal energy gap,
  and $\alpha$ the direction of current injection with respect to the antinodal direction in $ab$-plane contacts.
  }
  \label{tab:PCARS}
\end{table}

In both Figures\,\ref{fig:PCARS_fits}a and \ref{fig:PCARS_fits}b, the experimental spectra also display structures (that fall approximately between 7.5 and 15\,meV, as highlighted by the grey regions) that are perfectly consistent with those observed by STS but whose origin is not clear. As a matter of fact, it is well known that in all the Fe-based compounds, in addition to the structures associated to the energy gap(s), the point-contact Andreev-reflection spectra present higher-energy structures that are produced by  the strong coupling between carriers and the bosonic mode responsible for the pairing\,\cite{Tortello2010,Daghero2011,Daghero2014}. Indeed, if one uses the Eliashberg theory\,\cite{Eliashberg} to calculate the energy-dependent order parameters and then puts them in the Tanaka-Kashiwaya model for Andreev reflection\,\cite{KashiwayaPRB1996}, one obtains that these structures occur at an energy which is approximately $E_p = \Omega_b + \Delta_{max}$ where $\Omega_b$ is the energy of the spin resonance (that corresponds to the characteristic energy of the spectrum of the pairing bosons) and $\Delta_{max}$ is the amplitude of the larger gap. There are no data about the spin resonance in Rb-12442 but we know that it falls at about 16\,meV in K-12442  and at 15\,meV in Cs-12442, obeying the phenomenological law $\Omega_b = 5.8 k\ped{B} T_c$\,\cite{Hong2020PRL}. If this is true, in undoped K-12442 where $T_c \simeq 30$\,K one should have $\Omega_b \simeq 15$\,meV, and the electron-boson structures should fall at an even higher energy, which is incompatible with the features showing up already at 10\,meV. This is one of the puzzles of 12442 superconductors that would deserve deeper investigation.
Independent of their actual nature, the additional structures that fall in the grey regions must be excluded from the fit because they cannot be reproduced by the BCS-based models for Andreev reflection we are using here.

In the 5\%\,Ni-doped crystals, whose critical temperature is reduced by almost 10\,K with respect to the undoped ones, the shape of the spectra remains consistent. {\color{blue} Two examples of such spectra, for an $ab$-plane and a $c$-axis contact, are shown in Figure\,\ref{fig:PCARS_fits}c and \ref{fig:PCARS_fits}d, respectively. The experimental spectra (symbols)} display zero-bias peaks or cusps that, again, suggest the existence of  a nodal gap that we can represent as a $d$-wave one. The additional, higher-energy structures persist as well, but now they occur between 5 and 12.5\,meV and thus fall very close to the edge of the large gap, which makes it more difficult to disentangle the relevant contribution. Note that, again, the position of these structures is incompatible with a spin resonance energy that, according to the phenomenological law, should be at $\Omega_b = 5.8 k\ped{B} T_c = 10$\,meV, if the position of the peaks $E_p$ has to be equal to $\Omega_b + \Delta_{max}$.
In the $ab$-plane spectrum of Figure\,\ref{fig:PCARS_fits}c, the high zero-bias peak can be ascribed to the current being injected along the nodal direction ($\alpha=\uppi/4$), which maximizes the constructive interference between HLQ and ELQ. In the $c$-axis spectrum of Figure\,\ref{fig:PCARS_fits}d, instead, the zero-bias peak can be ascribed to the existence of zero-energy quasiparticle states at the gap nodes. {\color{blue} The fitting parameters are reported in Table\,\ref{tab:PCARS}. The broadening parameters remain smaller than the corresponding gap amplitude; at variance with the undoped crystal, the weight of the $d$-wave gap is smaller for $ab$-plane contacts than for $c$-axis ones. In the $c$-axis case, the barrier parameter $Z_d$ looks very small, but this is due to the aforementioned "$Z$-enhancing effect" of the different angular integration\,\cite{Daghero2011}.}

All curves for undoped and 5\%\,Ni-doped crystals, both in $ab$-plane and $c$-axis configurations can be fairly well fitted, in their central part, by an $s{-}d$ model, although the contribution of the $s$ gap is often barely visible. One can actually note that in \emph{all} cases (and not only in the examples shown in Figure\,\ref{fig:PCARS_fits}) the amplitude of both the gaps is systematically smaller in $c$-axis contacts than in $ab$-plane contacts. Owing to the 2D nature of the FS, it is difficult to ascribe this difference to an out-of-plane anisotropy of the gaps; rather, this might be an artifact due to the fact that the models used for $ab$-plane and $c$-axis contacts have the same physical meaning, but involve very different angular integrations. This difficulty could be overcome only by using a generalized 3D model that takes into account the real shape of the FS\,\cite{Daghero2011}.

As already pointed out, the $s{-}d$ model could mimic a nodal $s_\pm$ gap structure, with accidental nodes on some FS (Figure\,\ref{fig:gap_symmetries}b), or a $d_{x^2-y^2}$ symmetry with nodes only on the hole-like FSs (Figure\,\ref{fig:gap_symmetries}c). If ARPES measurements are right, the small gap (that is always seen as the nodal one, not only in our point-contact measurements but also in $\upmu$SR spectroscopy) does \emph{not} reside on the electron-like FSs but rather on the outer hole-like FS $\gamma$. 
Moreover, the spectral weight of the electron pockets is very likely to be vanishing in point-contact measurements as well as in STS measurements. Therefore, the observed PCS spectra could indicate that accidental nodes open up on the outer hole-like FS, which carries the smaller gap, within a nodal $s_\pm$ symmetry. 
The fact that the evidences of nodes persist upon Ni doping, however, raises some doubts about their accidental nature. Indeed, the Ni content $x = 0.05$ is very close to the compensation condition where $R\ped{H} =0$; therefore, such doping gives rise to a sensible reduction (increase) in the size of the hole-like (electron-like) FS sheets, thus pushing the system toward ``milder'' conditions. At the same time, Ni substitution introduces disorder, and this often results in a disappearance of the accidental nodes\,\cite{Mishra2009prb}. Here, the Ni substitution seems not to affect either the number of the gaps, or the presence of nodes, despite the clear effect on the size of the FSs.

If nodes are symmetry-protected, and not accidental, the symmetry must be a $d$-wave one, with lines of nodes intersecting each other in the $\Gamma$ point and crossing all the hole-like FSs (and, if the symmetry is $d_{xy}$, also the electron-like ones, as shown in Figure\,\ref{fig:gap_symmetries}d). But since in these compounds the hole-like FS sheets host the entire range of gap amplitudes (from 2\,meV at the $\gamma$ surfaces, to about 7\,meV at the $\beta_2$ surface) this symmetry would necessarily imply  that \emph{all} the gaps are nodal (maybe apart from the one residing on the electron-like FS, which is probably difficult to detect).

\begin{figure}
    \centering
    \includegraphics[width=\linewidth]{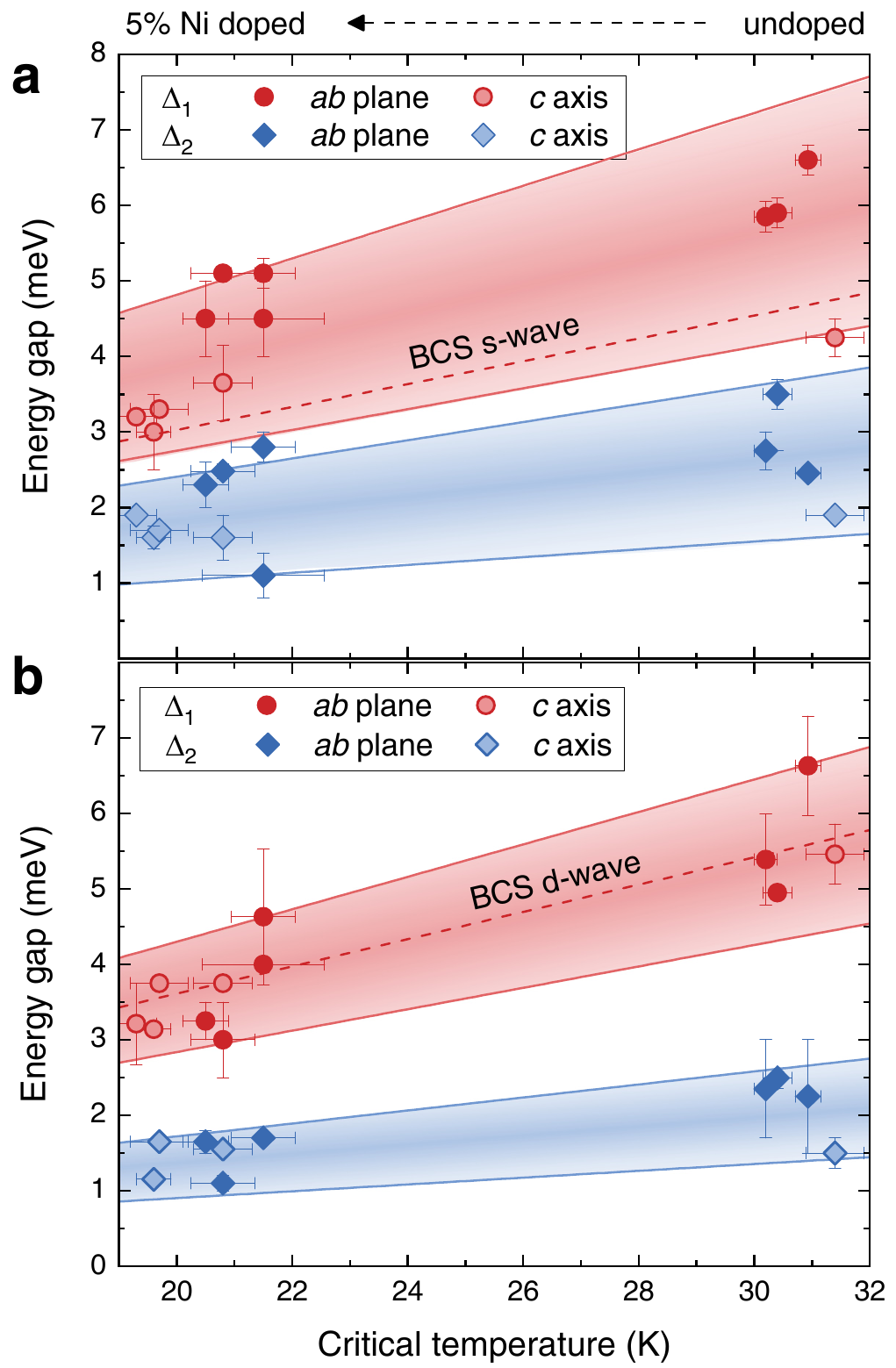}
    \caption{ 
        Amplitude of the superconducting energy gaps obtained from the fits of the experimental PCARS spectra vs the Andreev critical temperature of each point contact, using either the $s{-}d$ model (a) or the $d{-}d$ model (b) for the symmetry of the energy gap. The colored regions are bound by lines of constant gap ratio $2\Delta/k\ped{B}T_c$. The dashed lines are the gap amplitudes expected for single BCS gaps with $s$-wave (a) or $d$-wave (b) symmetry.
        Adapted from Ref.\,\onlinecite{Torsello2022npjQM}.
        }
    \label{fig:PCARS_summary}
\end{figure}

We thus tried to fit the PCARS spectra with a $d{-}d$ model, i.e. by using two $d$-wave gaps with the nodes in the same direction. In practice, this translates into using the same angle $\alpha$ (between the normal to the interface and the antinodal direction) for both gaps. The best-fitting curves to the experimental data of Figure\,\ref{fig:PCARS_fits} are represented by dotted blue lines. Clearly, these curves are hard to distinguish from the solid red ones, meaning that the $d{-}d$ fit works as well as the $s{-}d$ one. The fitting parameters are reported in Table\,\ref{tab:PCARS}.
{\color{blue} Here, the weight indicated refers to the small $d$-wave gap, and its value is systematically smaller in the $ab$-plane case than in the $c$-axis one in both the undoped and the 5\%\,Ni-doped crystals.
In $c$-axis contacts, the values of the barrier parameters are very small, and in the last case practically zero (they are not exactly zero to avoid divergences and numerical problems) but, as usual, this can be an artifact due to the angular integration\,\cite{YamashiroPRB1997}. Overall, the parameters of the $d-d$ fit vary in a more systematic way than those of the $s-d$ fit on going from the pristine to the doped sample, and from $ab$-plane to $c$-axis contacts. }

Figure\,\ref{fig:PCARS_summary} shows the amplitudes of the gaps obtained by using the $s{-}d$ model (panel a) and the $d{-}d$ model (panel b), as a function of the critical temperature of the contact. In the case of the $s{-}d$ model (Figure\,\ref{fig:PCARS_summary}a), the data are rather scattered and, although the small and the large gap are always clearly distinct in a single spectrum, the regions in which they fall (highlighted by colors) nearly overlap. The value of a BCS $s$-wave gap (i.e. $\Delta_s\apex{BCS}=1.76 k\ped{B} T_c$) is also shown by a dotted line. The values of the larger $s$-wave gap are generally (but not always) greater than the BCS one, while the amplitudes of the small $d$-wave gap are systematically smaller than that, in all contacts and for any $T_c$.  In the case of the $d{-}d$ model (Figure\,\ref{fig:PCARS_summary}b), the scattering of the data is much smaller, there is no clear dependence on the direction of current injection, and the regions in which the gap amplitudes fall are better separated. The dotted line, in this case, represents the value of a single, BCS $d$-wave gap (i.e., $\Delta_d\apex{BCS}=2.14 k\ped{B} T_c$\,\cite{Pakokthom1998JS}). 

\section{Coplanar waveguide resonator analysis}

Since the first evidence of nodes in the gap of 12442 compounds was provided by $\upmu$SR measurements of the superfluid density on polycrystalline samples\,\cite{Smidman2018PRB, Kirschner2018prb, Adroja2018JPSJ}, we decided to investigate this property also on the samples measured with PCARS. This is especially beneficial since the phase purity of our single-crystal samples allows us to exclude the possibility of contributions from residual nodal 122 phases that could exist in polycrystals.

\begin{figure}
    \centering
    \includegraphics[width=\linewidth]{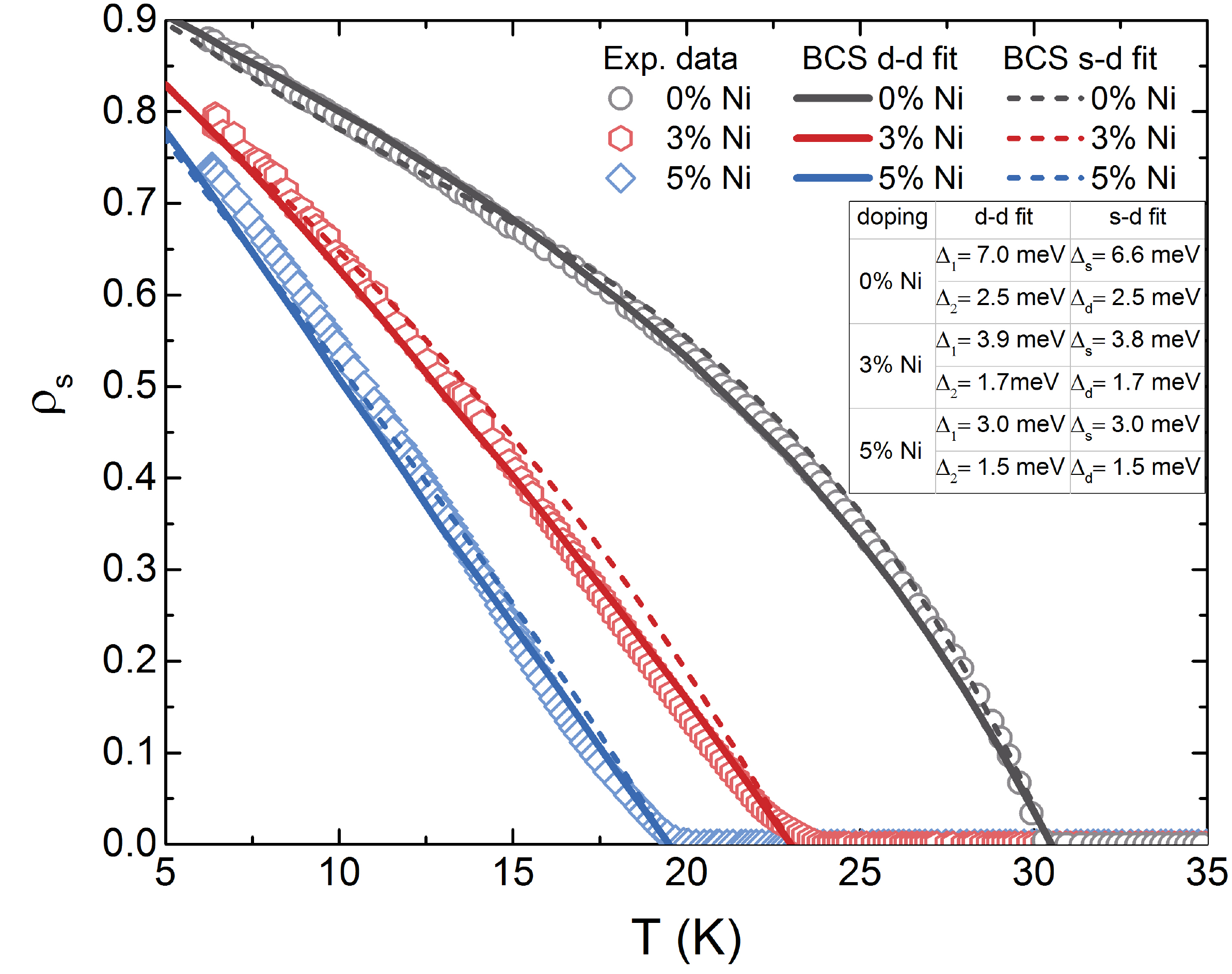}
    \caption{
    Temperature dependence of the normalized superﬂuid density  $\rho_S = \lambda(T=0)/\lambda(T)^2$ measured by CPWR (symbols) on single crystals with $x=0$ (gray circles), 0.03 (red hexagons) and 0.05 (blue diamonds) and their $d{-}d$ BCS ﬁts (solid lines with same color code) and $s{-}d$ fits (dashed lines). Adapted from Ref.\,\onlinecite{Torsello2022npjQM}.    
        }
    \label{fig:CPWR}
\end{figure}

We determined the superfluid density of our Rb-12442 single crystals via coplanar waveguide resonator (CPWR) measurements\,\cite{Torsello2019PRAppl, Ghigo2022Springer}, which is a technique ideally suited at analyzing small-size crystals where $\upmu$SR experiments would otherwise be impractical. In the CPWR approach, one measures the modifications to the resonance of a high-temperature superconducting resonator due to the presence of a small sample coupled to it. Exploiting a cavity perturbation approach, and after suitable calibration, it is then possible to extract the London penetration depth and hence the superfluid density of the investigated sample\,\cite{Torsello2019prb}. Similar to what was reported for the $\mu$SR measurements on the polycrystals, it is then possible to fit the experimental $\rho_S$ data with a suitable theoretical model\,\cite{Ghigo2018PRL, Torsello2019JOSC}.
Considering a two-gap BCS approach, one can write the superfluid density as\,\cite{Chandrasekhar1993}:
\begin{equation*}
    \rho_S(T)=\sum_i w_i \left[1+\frac{1}{\uppi}\int_0^{2\uppi}\int_{\Delta_i(\phi,T)}^{\infty} \frac{\partial f}{\partial E}\frac{E dE d\phi}{\sqrt{E^2-\Delta_i^2(\phi,T)}}\right]
\end{equation*}
where $i$ identifies the band, $\Delta_i(\phi,T)$ is the superconducting gap function, $f = [1 + \exp(E/k_\mathrm{B} T )]^{-1}$ is the Fermi function and $w_i$ is the mixing weight of the $i$-th gap contribution (constrained by $w_1+w_2=1$).
One can factorize the angular and temperature dependencies of the gap as $\Delta_i(\phi,T) =\Delta_{0,i} f_\phi(\phi) f_T(T)$, where the angular function for $d$-wave (nodal) superconductors is $f_\phi(\phi)=\cos(2\phi)$ and the temperature dependence can be approximated as $f_T(T)=\tanh{\left[1.82[1.018(T_\mathrm{c}/T - 1)]^{0.51} \right]}$.

\begin{figure*}
    \centering
    \includegraphics[width=\linewidth]{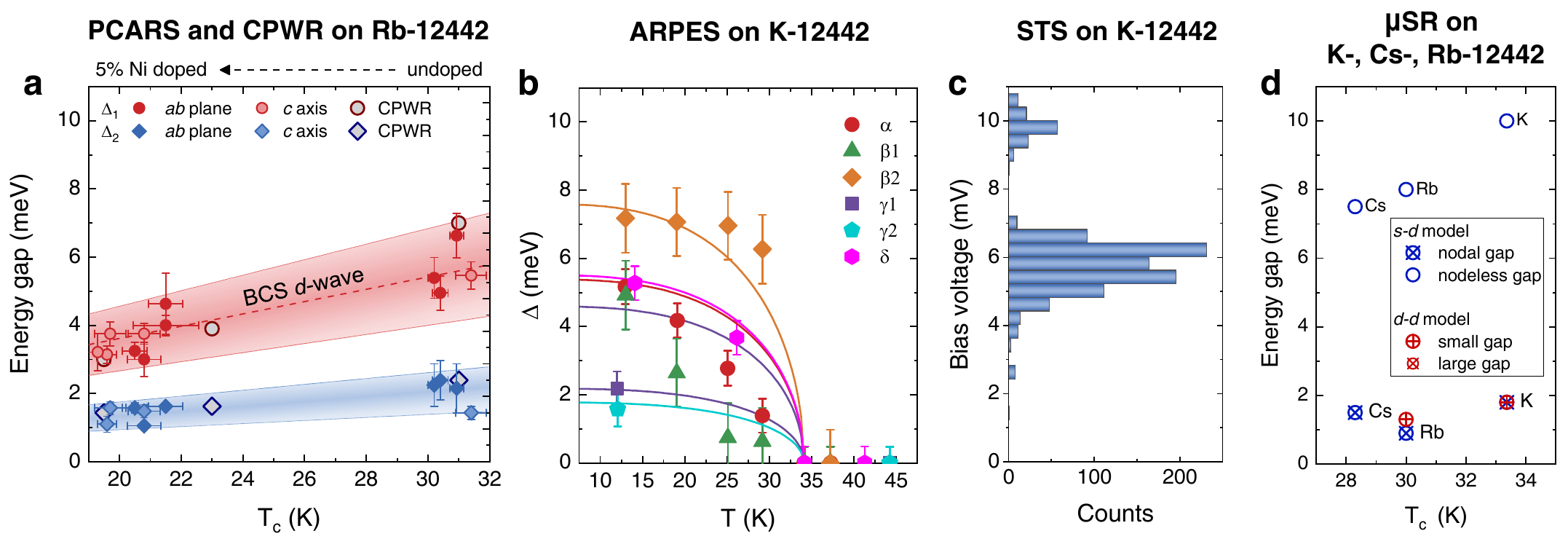}
    \caption{
    Comparison between the value of the superconducting energy gaps obtained via PCARS and CPWR on Rb-12442\,\cite{Torsello2022npjQM} (a), ARPES on K-12442\,\cite{Wu2020PRB} (b), STS on K-12442\,\cite{Duan2021PRB} (c), and $\upmu$SR on K-12442\,\cite{Smidman2018PRB}, Cs-12442\,\cite{Kirschner2018prb}, and Rb-12442\,\cite{Adroja2018JPSJ} (d).
    Panels (a-c) are adapted from Ref.\,\onlinecite{Torsello2022npjQM}, Ref.\,\onlinecite{Wu2020PRB} and Ref.\,\onlinecite{Duan2021PRB} respectively.
    }
    \label{fig:spectroscopy_comparison}
\end{figure*}

The superfluid density data and the BCS fits are shown in Figure\,\ref{fig:CPWR}. Our samples exhibit a linear temperature dependence of $\rho_S$ within the experimentally accessible temperature range, and the absence of kinks in the superfluid density curve. This latter feature is consistent with single-band superconductors and multigap systems with mainly interband coupling, such as the pnictides. When a multigap superconductor has predominant interband coupling, the superconducting gaps tend to show a gradual decrease and simultaneous closure at $T_c$ -- as evidenced by PCARS in our samples \cite{Torsello2022npjQM} -- and this results in a smooth superfluid density curve. Compared with the $\upmu$SR experiments on Rb-12442 polycrystals\,\cite{Adroja2018JPSJ}, our measurements display a less pronounced curvature at high $T$ and a wider $T$ range where the behavior of $\rho_S$ is quite linear, particularly in Ni-doped samples. These differences may be attributed to the type of samples (single vs polycrystalline). Nonetheless, the data from all doping levels of our samples series can be fitted with a BCS model involving two $d$-wave gaps, resulting in SC gap values that are consistent with those found by PCARS, as shown in Figure\,\ref{fig:spectroscopy_comparison}a). Similarly good fits can be obtained by fitting the data with an $s{-}d$ model, analogously to what was found in the $\upmu$SR experiments, possibly due to the contribution of several bands to the overall density. The gap values obtained by fitting the data are summarized in the table contained in Figure\,\ref{fig:CPWR}.

\section{Comparison between data from different spectroscopic techniques in 12442 compounds}

The results in literature are difficult to reconcile with one another, and the results of point-contact spectroscopy and CPWR measurements, unfortunately, do not solve the puzzle of the gap symmetry and structure, although the amplitudes of the gaps are in good agreement with those observed by ARPES and STS
. Figure\,\ref{fig:spectroscopy_comparison}a reports the $T_c$ dependence of the gaps obtained via PCARS and CPWR within the $d{-}d$ model, but similar ranges are obtained also with the $s{-}d$ model, as evidenced in Figure\,\ref{fig:PCARS_summary}c and Figure\,\ref{fig:CPWR}. 

A comparison with the results of ARPES\,\cite{Wu2020PRB} (Figure\,\ref{fig:spectroscopy_comparison}b) gives good results as far as the gap amplitudes are concerned. In particular, the range of ``large'' gaps determined by fitting the PCARS spectra and the superfluid density correspond very well with the range in which the gaps residing on the $\alpha$, $\beta$ and $\delta$ FSs are detected by ARPES. These gaps are very likely not to be distinguishable by point-contact spectroscopy especially because, in order to keep the number of fitting parameters within a reasonable limit, we cannot use more than two gaps at a time. Hence, the large gap we observe may actually be a sort of convolution of different gaps of similar amplitude, residing on different FSs. The gap on the $\gamma$ FSs corresponds rather well to the small gap we determined by PCARS and CPWR, irrespectively of the model used to fit the data ($s{-}d$ or $d{-}d$).

As far as STS is concerned, the distribution of gap values reported in Ref.\,\onlinecite{Duan2021PRB} (Figure\,\ref{fig:spectroscopy_comparison}c) agrees rather well with the range of the large gap we detect via PCARS and CPWR -- again, irrespectively of the specific model used. The smaller gap we measured, instead, corresponds to the position of the small peaks observed in some of the STS spectra and attributed to impurity states, rather than to a superconducting gap. Actually, the fact that STS is unable to detect the small gap, that according to ARPES resides on the largest hole-like FS, is rather puzzling. 

Conversely, the gap amplitudes determined by $\upmu$SR (Figure\,\ref{fig:spectroscopy_comparison}d) are in poor agreement with the results of all the more direct spectroscopic techniques. In particular, the $d{-}d$ fit of the superfluid density measured by $\upmu$SR gives amplitudes of the large gap that fall out of the vertical scale, being equal to 14\,meV in both K-12442\,\cite{Smidman2018PRB} and Rb-12442\,\cite{Adroja2018JPSJ}; the small gap is instead in better agreement with the one determined by the other techniques, being equal to 1.8\,meV in K-12442\,\cite{Smidman2018PRB} and 1.3\,meV in Rb-12442\,\cite{Adroja2018JPSJ}. Similar amplitudes of the nodal gap are given by the $s{-}d$ fit to the superfluid density, which gives values that are always smaller than 2\,meV (and precisely, 1.8\,meV in K-12442\,\cite{Smidman2018PRB}, 0.9\,meV in Rb-12442\,\cite{Adroja2018JPSJ} and 1.5\,meV in Cs-12442\,\cite{Kirschner2018prb}). Concerning the large gap, the agreement between $\upmu$SR and the other techniques is improved when the $s{-}d$ fit to the superfluid density is considered. Nevertheless, the nodeless gap remains by far too big with respect to the ones determined by ARPES: 10\,meV in K-12442\,\cite{Smidman2018PRB}, 8\,meV in Rb-12442\,\cite{Adroja2018JPSJ} and 7.5\,meV in Cs-12442\,\cite{Kirschner2018prb}. Since the fits to the superfluid density measured via CPWR -- using both $s{-}d$ and $d{-}d$ models -- give gap amplitudes in excellent agreement with those obtained via other spectroscopic techniques, this mismatch could be potentially ascribed to the polycrystalline nature of the samples probed via $\upmu$SR. This further highlights the importance of relying on samples with excellent phase purity when attempting an accurate quantitative investigation of the intrinsic properties of a given compound.

In contrast with the gap amplitudes, the evidence of nodal quasiparticle states in 100\% of the PCARS spectra are completely at odds with the results of both ARPES and STS in K-12442, while they are in agreement with the linear low-temperature dependence of the superfluid density as probed by both $\upmu$SR and CPWR. In principle, this discrepancy calls for further measurements of ARPES and STS in this specific compound and, in general, indicates the great complexity of these iron-based compounds. 

\section{Conclusions}

The recently-discovered 12442 compounds are one of the most peculiar and interesting families among iron-based superconductors. Their structure, made up of FeAs bilayers separated by insulating CaF\ped{2} layers, makes them highly anisotropic and similar, in many respects, to double-layer cuprates. Nonetheless, the extent to which these compounds can be considered as the iron-based counterpart of cuprates is still poorly understood, especially as it pertains to the symmetry of the superconducting order parameter.
In this work we have reviewed the structural, electronic and superconducting properties of this class of materials, with a specific focus on the structure and symmetry of the superconducting energy gap as probed by different spectroscopic techniques, including angle-resolved photoemission spectroscopy (ARPES), muon spin rotation ($\upmu$SR), scanning tunnelling spectroscopy (STS), point-contact Andreev reflection spectroscopy (PCARS) and coplanar-waveguide resonator (CPWR) measurements.

While nearly all experimental probes agree in detecting the presence of multiple superconducting gaps opening on the complex, multi-band Fermi surface of these compounds, the exact amplitudes pertaining nominally to the same gaps vary between the different techniques. In the undoped compounds, $\upmu$SR systematically observes the larger gap to lie at much higher (and possibly nonphysical) energies with respect to the other techniques ($4{-}7$\,meV); conversely, STS ascribes the spectral features observed at about $\sim2$\,meV (where other probes detect the opening of the smaller gap) to the presence of defect states. Additionally, both PCARS and STS observe the presence of spectral features around 10\,meV -- not detected by other techniques -- which remain of unknown origin, being hard to reconcile either with the presence of an even larger gap or to electron-boson coupling.
The starkest disagreement is however found in the symmetry of the order parameter, with neutron scattering, optical conductivity, thermal conductivity, STS, and ARPES measurements being most compatible with nodeless energy gap(s), while $\upmu$SR, specific heat, PCARS and CPWR all exhibit unambiguous evidence for the existence of quasiparticle nodal lines on at least one, small energy gap. Furthermore, PCARS and CPWR show these nodal lines to be surprisingly robust against Ni doping, hinting at the fact that -- at least in the Rb-12442 compound -- these might be symmetry-imposed (i.e., associated with a $d$-wave symmetry of the order parameter) rather than accidental (i.e., associated with the more standard nodal $s_\pm$ symmetry).

Since, up to now, the aforementioned spectroscopic techniques have been applied to different members of the 12442 family, this complicated situation calls for further and more systematic investigations -- by using complementary experimental tools on single crystals of the same compound -- and possibly for quasiparticle interference experiments, that are considered to allow a robust determination of the gap structure in iron-based superconductors\,\cite{Hirschfeld2015prb}.


\section*{Acknowledgments}
This work was supported by the Italian Ministry of Education, University, and Research through the PRIN-2017 program: 
E.P. and D.D. acknowledge support from project “Quantum2D”, Grant No. 2017Z8TS5B;
D.T. and G.G. acknowledge support from project “HIBiSCUS”, Grant No. 201785KWLE. 
D.T. also acknowledges support by the “Programma Operativo Nazionale (PON) Ricerca e Innovazione 2014--2020”.

%

\end{document}